\documentclass[prb,twocolumn,unsortedaddress,showpacs,floatfix]{revtex4}

\usepackage{graphicx}
\usepackage{array}
\usepackage{amsmath,amsfonts,amssymb}

\input{tcilatex}

\begin{document}

\title{Susceptibility and dilution effects of the kagom\'{e}  bi-layer geometrically
frustrated network. A Ga-NMR study of SrCr$_{9p}$Ga$_{12-9p}$O$_{19}$.}

\author{L. Limot}
\altaffiliation[Present address: ]{Institut f\"{u}r Experimentelle und Angewandte Physik, Christian-Albrechts-Universit\"{a}t zu Kiel, D-24098 Kiel, Germany}
\author{P. Mendels}
\affiliation{Laboratoire de Physique des Solides, UMR 8502, Universit\'e Paris-Sud, F-91405 Orsay, France}
\author{G. Collin}
\affiliation{Laboratoire L\'eon Brillouin, CE Saclay, CEA-CNRS, F-91191 Gif-sur-Yvette, France}
\author{C. Mondelli}
\author{B. Ouladdiaf}
\author{H. Mutka}
\affiliation{Institut Laue-Langevin, B.P. 156, F-38042 Grenoble Cedex 9, France}
\author{N. Blanchard}
\affiliation{Laboratoire de Physique des Solides, UMR 8502, Universit\'e Paris-Sud, F-91405 Orsay, France}
\author{M. Mekata}
\affiliation{Department of Applied Physics, Fukui University, Fukui 910, Japan}

\date{\today}

\begin{abstract}
The archetype of geometrically frustrated compounds SrCr$_{9p}$Ga$_{12-9p}$O$%
_{19}$ is a kagom\'{e} bi-layer of Heisenberg Cr$^{3+}$ ions ($S=3/2$) with
antiferromagnetic interactions. We present an extensive gallium NMR study
over a broad Cr-concentration range ($0.72\leq p\leq 0.95$). This allows us to
probe locally the susceptibility of the kagom\'{e} bi-layer and separate the
intrinsic properties due to geometric frustration from those related to
site dilution. Compared to the partial study on one sample, $p=0.90$, presented
in \prl\textbf{85}, 3496 (2000), we perform here a refined study of the
evolution of all the magnetic properties with dilution, with a large
emphasis on the lowest diluted $p=0.95$ sample synthesized for this study.

Our major findings are: 1) The intrinsic kagom\'{e} bi-layer susceptibility
reaches a maximum at a temperature of $\approx 40-50$ K, which we show here to
be robust to a dilution as high as $\approx 20\%$.
This maximum is the signature of the
development of short range antiferromagnetic correlations in the kagom\'{e} bi-layer ; 2) At low-$T$, a
highly dynamical state induces a strong wipe-out of the NMR intensity,
regardless of dilution; 3) The low-$T$ upturn of the macroscopic
susceptibility is associated to paramagnetic defects which stem from the
dilution of the kagom\'{e} bi-layer. The low-$T$ analysis of the $p=0.95$ NMR
lineshape coupled with a more accurate determination of the nuclear
hamiltonian at high-$T$, allows us to discuss in detail the nature of the defect.
Our analysis suggests that the defect can be associated with a staggered
spin-response to the vacancies of the kagom\'{e} bi-layer. This,
altogether with the maximum in the kagom\'{e} bi-layer susceptibility, is
very similar to what is observed in most low-dimensional antiferromagnetic
correlated systems, even those with a short spin-spin correlation length; 4) The
spin glass-like freezing observed at $T_{g}$ $\approx 2-4$ K is not driven by
the dilution-induced defects.
\end{abstract}


\maketitle

\altaffiliation[Present address: ]{Institut f\"{u}r Experimentelle und
Angewandte Physik, Christian-Albrechts-Universit\"{a}t zu Kiel, D-24098
Kiel, Germany}

\affiliation{Laboratoire de Physique des Solides, UMR 8502, Universit\'e
Paris-Sud, F-91405 Orsay, France}

\affiliation{Laboratoire L\'eon Brillouin, CE Saclay, CEA-CNRS, F-91191
Gif-sur-Yvette, France}

\affiliation{Institut Laue-Langevin, B.P. 156, F-38042 Grenoble Cedex 9,
France}

\affiliation{Laboratoire de Physique des Solides, UMR 8502, Universit\'e
Paris-Sud, F-91405 Orsay, France}

\affiliation{Department of Applied Physics, Fukui University, Fukui 910,
Japan}

\section{Introduction}
\label{intro}

Under certain circumstances, it is impossible for magnetic systems to
minimize simultaneously all the interactions between the spins. The system is then
frustrated. This is known to occur in the spin glass (SG) compounds where
the disorder on the magnetic network induces a competition between the interactions.
In a vast variety of systems, the frustration can arise from the geometry of the
lattice itself, without disorder,\cite{Liebmann} as in the case of
the triangular-based antiferromagnetic (AF) networks. The kagom\'{e} ($d=2$) and
the pyrochlore ($d=3$) networks with AF interactions are a particular class of
geometrically frustrated networks where the triangles (for the kagom\'{e}
lattice) and the tetrahedras (for the pyrochlore lattice) share corners instead
of sides as for the familiar triangular network.

Within a classical theory, the ground state is built on triangles and/or
tetrahedras with a zero total magnetic moment. Remarkably, the corner-sharing
geometry gives rise to a macroscopic degeneracy of the ground state.\cite
{Degeneracy1,Degeneracy2} By macroscopic, we mean that the energy of the kagom\'{e}
or of the pyrochlore ground state is invariant under a rotation of a finite
number of spins. Through these rotations, the system can explore all the
spin configurations making up the ground energy level. The energy spectrum
is therefore characterized by the existence of zero energy excitations, the
so-called soft modes.\cite{Zeromodes} In particular, the soft modes associated
to a small number of spins are extremely efficient in destroying a long range
magnetic order. A non magnetic ground state is predicted at $T\rightarrow 0$ K, with a
spin-spin correlation function $\left\langle S_{i}S_{j}\right\rangle \sim
exp(-r_{ij}/\xi )$, where the correlation length $\xi $ does not exceed
twice the lattice parameter.\cite{Correlation} The huge reservoir of soft
modes leads also to a low energy shift of the excitation spectrum, so that
an unusually highly dynamical state is predicted at low-$T$.\cite{Dynamics}
This peculiar ground state is commonly called a cooperative paramagnet state
or a ``spin liquid state''.

Quite remarkably, a different description of quantum nature for $S=1/2$
geometrically frustrated AF networks yields similar results. As first
elaborated by Anderson,\cite{Anderson} and validated through numerical
calculations, the ground state can be constructed on spin singlet pairs and
not on non magnetic triangles (or tetrahedras).\cite{Quantum} In particular,
numerical studies point at the existence of a ``gap'' in the magnetic
excitation spectrum of the kagom\'{e} $S=1/2$ network,\cite{Waldtmann} and
maybe in the pyrochlore $S=1/2$ network.\cite{Canals} The term gap is
actually inappropriate as a continuum of singlet states is imbedded between
the singlet ground state and the first excited triplet state. These
singlet excitations are in some regards the analog of the classical soft
modes. A spin liquid ground state is again favored, with unusual thermodynamic
properties such as a high entropy at low-$T$.\cite{Lhuillier}

The discovery of the kagom\'{e}-based insulator SrCr$_{9p}$Ga$_{12-9p}$O$%
_{19}$ in 1988,\cite{Obradors} brought considerable attention on geometric
frustration. Since then, an intense mapping of the low-$T$ physics of the
kagom\'{e} and of the pyrochlore compounds has been carried out.\cite
{ReviewExp,Gingras} Original properties have been uncovered, encompassing
anomalous SG states, \cite{SG} and ice-like ground states.\cite{Ice}
However, nowadays, only a few compounds are good candidates for a spin
liquid ground state. Following Villain's early work,\cite{Villain} the
discrepancy between experience and theory is to be found in the perturbations
to the ideal AF Heisenberg hamiltonian with nearest neighbor spin-spin
interactions. The dilution of the network,\cite{Dilution,Degeneracy2}
the interactions other than nearest neighbor,\cite{Non-NNN} the dipolar
interaction,\cite{Dipolar} the anisotropy,\cite{Anisotropy} etc. are
perturbations which differentiate the geometrically frustrated compounds and
make them deviate from the ideal spin liquid behavior. Each deviation can
potentially induce a long range order, usually quite complex. In all
experimental studies, one must therefore discriminate between the properties
related to geometric frustration and those related to these perturbations, that we
may label by the general term of disorder.

\begin{figure}[tbp]
\includegraphics[width=6cm,bbllx=60,bblly=245,bburx=545,bbury=750,clip=]{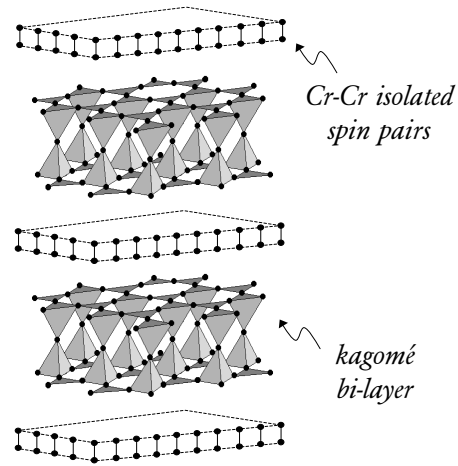}
\caption{The magnetic lattice of SCGO is a stacking of kagom\'e bi-layers of
Cr$^{3+}$ ions ($\approx 78\%$ of the Cr sites), separated by Cr-Cr isolated
spin pairs ($\approx 22\%$ of the Cr sites).}
\label{fig:1}
\end{figure}

However, even in the presence of these limiting parameters, spin liquid-like
compounds do exist.\cite{SL} More than in any other compound, spin liquid-like
properties are observed in SrCr$_{9p}$Ga$_{12-9p}$O$_{19}$ $\left[ \text{SCGO}%
(p),0\leq p<1\right]$. The geometric frustration in SCGO arises from a kagom\'{e}
bi-layer of Heisenberg Cr$^{3+}$ ions ($S$=3/2), a quasi$-d=2$ network of
two kagom\'{e} layers connected by a triangular lattice linking layer (Fig.~%
\ref{fig:1}). The disorder in SCGO stems from the dilution of the  Cr-network
by non magnetic Ga$^{3+}$ ions. To date, all SCGO crystals reported in
literature are non stoichiometric ($p<1$).

Although an anomaly is observed at low-$T$ ($T_{g}\approx 2-4$ K for $%
0.6\leq p<1$) in the macroscopic susceptibility suggesting the occurrence of
a SG state,\cite{RamirezTg,Martinez1,Martinez2} all the other experimental data
available on SCGO point at the existence of a spin liquid-like ground state. The
neutron diffraction pattern at $T<T_{g}$ is characterized by a broad peak,
from which is extracted a spin-spin correlation length of twice the Cr-Cr
distance ($\xi \approx 2d_{Cr-Cr}$).\cite{Broholm} A more refined neutron
experiment showed that the diffraction pattern is in agreement with the
existence of sub-groups of spins (singlets, triangles or tetrahedras) of
zero magnetic moment.\cite{LeeShort} Neutrons also revealed that only a
fraction of the Cr$^{3+}$ moment is frozen and $\mu $SR experiments revealed
the existence of a strongly fluctuating ground state,\cite{Uemura,AmitMuSR}
a picture further supported by recent specific heat measurements that
indicate that only $\sim 50\%$ of the total entropy is removed below 100 K.
\cite{RamirezCv}

From the macroscopic susceptibility ($\chi _{macro}$), one can extract that
a strong AF interaction couples the neighboring spins with a characteristic
Curie-Weiss temperature $\Theta _{macro}\approx 500-600$ K. The Curie-Weiss
behavior of $\chi _{macro}$ extends to temperatures $T\ll \Theta _{macro}$,
which is recognized as the most typical signature of frustration.\cite{ReviewExp,Gingras}
Puzzlingly, at very low-$T$, $\chi _{macro}$ progressively deviates from
Curie-Weiss behavior and exhibits a behavior closer to a simple Curie law.
This property of SCGO is actually encountered in the majority of the
geometrically frustrated magnets. This is in deep contrast with the
non magnetic ground state predicted for a spin liquid. It was conjectured
that this Curie upturn might originate from the dilution, i.e. may not be an
intrinsic property of the kagom\'{e} bi-layer susceptibility.\cite{Schiffer}
If this is indeed the case, what is then the susceptibility of the
kagom\'{e} bi-layer? And what is the underlying mechanism which triggers the
Curie upturn?

\begin{figure}[t]
\includegraphics[width=8.6cm,bbllx=20,bblly=310,bburx=590,bbury=715,clip=]{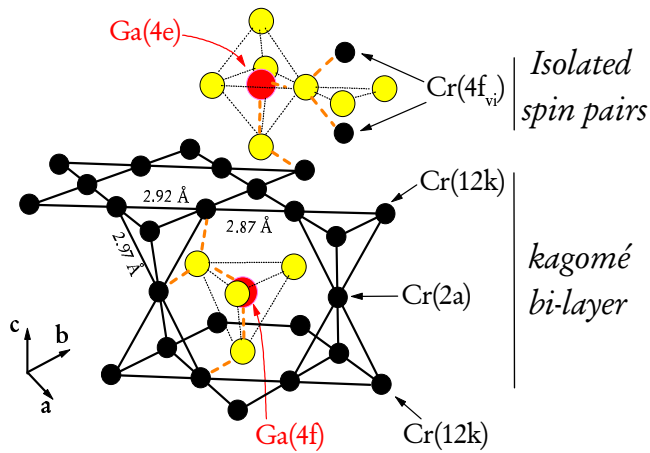}
\caption{The crystal structure of ideal SrCr$_{9}$Ga$_{3}$O$_{19}$. The
light grey circles represent the oxygen atoms (we have restrained their
number for clarity). The Sr atoms are not represented. The thick dashed lines
show the typical hyperfine coupling paths of the gallium nuclei to various Cr sites
through the oxygen ions.}
\label{fig:2}
\end{figure}

These are the main topics addressed in this work through the NMR of gallium
nuclei of SCGO. The $^{69}$Ga and $^{71}$Ga nuclei ($I=3/2$) are local probes
coupled to the Cr$^{3+}$ ions. The gallium nuclei labeled Ga($4f$) are at the
heart of frustrated physics, as they are exclusively coupled to the Cr$^{3+}$
ions of the kagom\'{e} bi-layer (Fig.~\ref{fig:2}). The present NMR study
was carried out on different Cr-concentrations of SCGO ($0.72\leq p\leq 0.95$%
). The comparative study of the samples shows that Ga($4f$)-NMR can discern
between the geometric frustration related properties and the disorder related properties
of SCGO: we are able to probe independently the kagom\'{e} bi-layer susceptibility and
the effects related to dilution, and thus answer both questions on a firm
experimental ground.

The general outline of this paper is the following. After a detailed
presentation of the samples used in this work, their characterization and
a description of their magnetic couplings (Sec.~\ref{section:samples}), we focus on the
specificities of the Ga-NMR spectrum of SCGO: the gallium sites in SCGO, the nuclear
parameters of the Ga($4f$) resonance line and the optimization of its detection
(Sec.~\ref{Ga-NMR spectrum}), with a special emphasis on the new
features brought by the study of the least diluted sample $p=0.95$. We also
show that NMR is a very refined technique for characterizing the amount of
Ga/Cr substitution, especially in the low dilution limit. The following
sections address the main points relevant to the frustrated physics.
Compared to Ref.~\onlinecite{Philippe}, the study of the dilution effects is
entirely new. After a brief overview of the raw NMR spectra and some
experimental details (Sec.~\ref{Dilution and $T$-dependence of the Ga($4f$)
spectrum}), we present in Sec.~\ref{section:kagome bi-layer susceptibility}
the first part of our comparative study. By probing the susceptibility on a
local stand through the NMR shift, we identify the intrinsic, i.e.
dilution-independent, susceptibility typical of the SCGO family $-$ the
archetype of geometrically frustrated compounds. We then discuss our
experimental results in the light of the existing models and calculations. The
main topic of Sec.~\ref{section:dilution effects} is the study of the
dilution effects. Using the NMR width, we can track down in SCGO the effect
associated with the dilution of the magnetic network and isolate its
contribution to the macroscopic susceptibility. Moreover, we demonstrate
that the impact of the site dilution is not an effect simply localized around the substitution
site. As we discuss, this sets specific constraints on the theoretical models
describing the dilution effects. We finally suggest an interpretation of our NMR
results based on the general context of the AF correlated systems. A summary and
the concluding remarks can be found in Sec.~\ref{Conclusion}.

\section{Samples and magnetic couplings}

\label{section:samples}

The Cr$^{3+}$ ions of SCGO occupy three distinct sites which are denoted $%
12k $, $2a$ and $4f_{vi}$ (Fig.~\ref{fig:2}). The Cr($12k$) sites represent $%
2/3$ of the total number of Cr$^{3+}$ ions and are arranged to form a
kagom\'{e} lattice. The two kagom\'{e} planes are separated by a Cr($2a$)
triangular layer ($1/9$ of the Cr$^{3+}$ ions). The remaining $2/9$ of the Cr%
$^{3+}$ ions occupy the Cr($4f_{vi}$) sites. A realistic description of the
Cr-Cr magnetic couplings in SCGO showed that this compound has couplings
similar to Cr$_{2}$O$_{3}$.\cite{Lee} As was evidenced in Ref.~%
\onlinecite{Lee}, SCGO is made up of only two magnetic entities. 1) The
kagom\'{e} bi-layer, i.e. the kagom\'{e}-Cr(2a)-kagom\'{e} structure,
with an average AF coupling of $J_{bi-layer}\approx 80$ K as we establish in
Sec.~\ref{section:kagome bi-layer susceptibility}. 2) The Cr($4f_{vi}$)-Cr($%
4f_{vi}$) spin pairs with an AF coupling of $J_{pair}=216(2)$ K, each pair
being isolated from the others. The full structure is obtained by the
stacking: spin pairs/kagom\'{e} bi-layer/spin pairs/kagom\'{e} bi-layer etc.
The interaction between the kagom\'e bi-layer and the spin pairs is small ($\sim 1$ K). Since all
the Cr($4f_{vi}$) spin-pairs form non magnetic singlets at low-$T$ ($T\ll
J_{pair}$, see Appendix), the low-$T$ properties of SCGO are expected to
reflect those of the kagom\'{e} bi-layer network only.

\begin{table}[b]
\caption{The characteristic susceptibility parameters for some of the SCGO samples studied. $\protect%
\mu _{eff} (\protect\mu _{B})$ and $\Theta _{macro}$ are extracted by
fitting the high-$T$ behavior of $\protect\chi _{macro}^{-1}$ ($T\geq 150$
K) to a Curie-Weiss law $\propto \left[\protect\mu _{eff}^{2}/(T+\Theta
_{macro})\right]^{-1}$.}
\label{tab:1}%
\begin{ruledtabular}
\begin{tabular}{@{\extracolsep{0ptplus1fil}}cccc}
$p$    & $\mu _{eff} (\mu _{B})$ & $\Theta _{macro}$ (K) & $T_{g}$ (K) \\
\colrule
0.72    & 3.85(5) & 356(11) & 2.3 \\
0.81    & 4.00(2) & 439(7)  & 3.3 \\
0.89    & 4.05(2) & 501(7)  & 3.5 \\
0.90    & 4.21(1) & 560(5)  & 3.2 \\
0.95    & 4.23(2) & 608(7)  & 3.6 \\
\end{tabular}
\end{ruledtabular}
\end{table}

From the previous considerations, it is quite clear that the kagom\'{e}
bi-layer structure is more complex to model than a pure kagom\'{e} system.
However, SCGO is ideal in many regards. As mentioned in the introduction,
the interactions in the magnetic hamiltonian other than a nearest neighbor
interaction, are likely to modify the spin liquid nature of the ground state.
These interactions are here extremely small compared to the Cr-Cr
interaction in the kagom\'{e} bi-layer. The anisotropy of Cr$^{3+}$ is $%
\approx 0.08$ K (spins are therefore Heisenberg), \cite{ESR} the dipolar
interaction is $\sim 0.1$ K, and the next nearest neighbor interactions $J_{\text{n.n.n}}$ are
most likely $\left|J_{\text{n.n.n}}\right|<4$ K, as in Cr$_{2}$O$_{3}$. \cite{Samuelsen} These
microscopical details might not be totally uncorrelated to the spin liquid-like
behavior observed in SCGO.

\begin{figure}[b]
\includegraphics[width=7cm,bbllx=25,bblly=145,bburx=590,bbury=600,clip=]{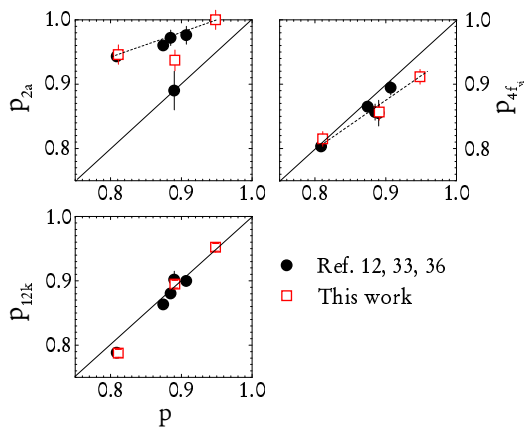}
\caption{The relative Cr-concentration on the $12k$, $2a$ and $4f_{vi}$ sites
determined by the neutron measurements ($p_{12k},p_{2a},p_{4f_{vi}}$) versus the
overall Cr-concentration ($p$) of our $p=0.81,0.89,0.95$ samples (open
squares) and of other samples in the literature (closed circles). The
Cr-concentration of the three sites follows stoichiometry when the symbols
fall on the solid line.}
\label{fig:3}
\end{figure}

The present study was performed on a series of seven Cr-contents,
corresponding to concentrations of $p=0.72,0.81,0.89,0.90,0.91,0.93,0.95$. All the
samples are ceramics and were synthesized by a solid state reaction of SrCO$%
_{3}$, Cr$_{2}$O$_{3}$ and Ga$_{2}$O$_{3}$ in air at $1350^{\circ }$ C,
typical for SCGO. The reaction products were checked by X-ray diffraction and by
macroscopic susceptibility measurements. The physical parameters for $\chi
_{macro}$ yielded results in agreement with literature (Tab.~\ref{tab:1}).
More refined characterizations were also performed. The $p=0.81,0.89,0.95$
samples were investigated by high resolution neutron diffraction with the
D2B spectrometer of the Insitut Laue-Langevinin in order to estimate the
occupation of the three Cr sites of SCGO, respectively $p_{12k}$, $p_{2a}$
and $p_{4f_{vi}}$, otherwise non measurable by X-ray diffraction. The results
are presented in Fig.~\ref{fig:3}. The $p=0.72$ and $0.90$ samples were the
object of $\mu $SR studies.\cite{Uemura,AmitMuSR} Finally, the Cr-concentration
of the samples was also checked directly by Ga-NMR (detailed in Sec.~\ref
{Ga-NMR spectrum}).

The synthesis of samples with higher Cr-content than $p=0.95$ failed. The X-ray
diffraction showed a Cr$_{2}$O$_{3}$ parasitical presence of $\agt$ $0.5\%$
(our detection threshold) in the final product of $p=0.95$ and all the
additional Cr$_{2}$O$_{3}$ introduced in the solid state reaction to reach
higher concentrations than $p=0.95$ simply did not react. This sets the limit
of this synthesis method to the $p=0.95$ concentration. Interestingly, this
limit may be related to the fact that the Ga/Cr substitution in SCGO is not
uniform, i.e. the Cr-concentration is not equal to $p$ on all the sites.
As shown in Fig.~\ref{fig:3}, the Ga/Cr substitution on the Cr($12k$) sites
corresponds to stoichiometry ($p\approx p_{12k}$), whereas in contrast
the Cr($2a$) sites are robust to substitution ($p_{2a}>p$), the non-stoichiometric
gallium occupying preferentially the Cr($4f_{vi}$) sites ($p_{4f_{vi}}<p$).
We note that no gallium is present on the Cr($2a$) sites of the $p=0.95$ sample
and that concomitantly the parasitical Cr$_{2}$O$_{3}$ problem occurs at this
same Cr-concentration, an indication that the two phenomena may be related to the
same chemical constraint.

\begin{figure}[t]
\includegraphics[width=8cm,bbllx=40,bblly=355,bburx=505,bbury=615,clip=]{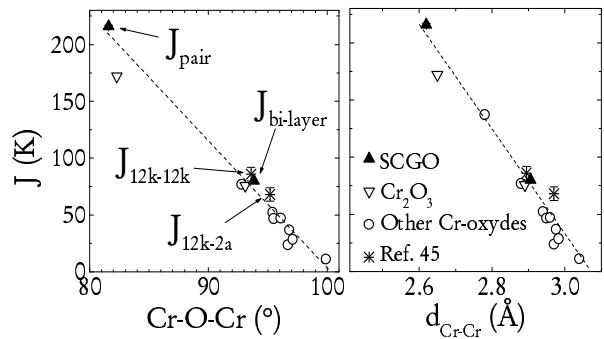}
\caption{The Cr-Cr couplings as a function of the Cr-O-Cr angle (\textit{left}) and
the Cr-Cr distance (\textit{right}) for: the average Cr-Cr coupling of the SCGO
kagom\'e bi-layer (see Sec.~\ref{Evidence for a two component macroscopic
susceptibility}), the SCGO Cr($4f_{vi}$)-Cr($4f_{vi}$) pairs,\protect\cite
{Lee} Cr$_{2}$O$_{3}$,\protect\cite{Samuelsen} and a series of Cr-oxydes.
\protect\cite{Motida} The stars stand for the Cr($12k$)-Cr($12k$) and the
Cr($12k$)-Cr($2a$) couplings of SCGO's kagom\'e bi-layer which we determined in
a previous publication via a refined analysis of the NMR shift.\protect\cite
{LaurentHFM2000}}
\label{fig:4}
\end{figure}

A last point we would like to address, is the influence of the Ga/Cr
substitution on the crystal parameters $a$ and $c$. The largest structural
change we observed is a $\approx 0.004$ {\AA} elongation of
the Cr($12k$)-Cr($2a$) distance in the $p=0.72$ sample ($2.975$ {\AA }) compared to
the $p=0.95$ sample ($2.971$ {\AA }). A study on a series of
chromium-based oxides with Cr-O-Cr bonding angles $\sim 90^{\circ }$ (as in
SCGO) showed that the Cr-Cr exchange constant decreases linearly with
increasing Cr-Cr distance (Fig.~\ref{fig:4}), with a slope of $\Delta
J/\Delta d_{Cr-Cr}\approx 450$ K/{\AA },\cite{Motida} from which we extract
a negligible variation of the SCGO exchange constants at most of $\Delta
J\approx 3$ K. We therefore expect the sizeable \textit{variation of the
magnetic properties of SCGO with }$p$ to reflect only the effect of the
lattice \textit{dilution}.

\section{Ga-NMR spectrum}

\label{Ga-NMR spectrum}

The present NMR study was performed on the $^{69}$Ga ($^{69}\gamma =10.219$ MHz/T,
$^{69}Q=0.178\times 10^{-24}$ cm$^{2}$) and the $^{71}$%
Ga ($^{71}\gamma =12.982$ MHz/T, $^{71}Q=0.112\times 10^{-24}$ cm%
$^{2}$) nuclei of SCGO powder samples ($\gamma $ and $Q$ are respectively
the gyromagnetic ratio and the quadrupolar moment of the nucleus). The gallium ions are
present on two distinct crystallographic sites, which are designated by Ga($%
4f$) and Ga($4e$) (Fig.~\ref{fig:2}). A previous spectral analysis carried
out on the $p=0.90$ sample in a sweep field set up at a radio-frequency of $%
\nu _{rf}=131$ MHz, successfully assigned each site to the corresponding
peak in the NMR spectrum.\cite{AmitNMR} There, it was shown that the Ga-NMR
spectrum of both isotopes is actually the sum of three contributions: Ga($4f$%
), Ga($4e$) and an extra contribution related to the presence of
non-stoichiometric gallium on the Cr sites, which we label by Ga($sub$).

As was pointed out in Ref.~\onlinecite{AmitNMR}, the interest of Ga-NMR
resides in the fact that the gallium nuclei are coupled to the neighboring
magnetic Cr$^{3+}$ ions through a Ga-O-Cr hyperfine bridge (Fig.~\ref{fig:2}).
In particular, the $^{69,71}$Ga($4f$) nuclei are exclusively coupled to
the kagom\'{e} bi-layer, 9 from the 2 kagom\'{e} adjacent Cr($12k$)-layers and
3 from the intermediate Cr($2a$)-sites. Through Ga(4f)-NMR we are then able to
probe locally the magnetic properties of the kagom\'{e} bi-layer. The present NMR
study is therefore devoted to Ga($4f$)-NMR.

Along with the hyperfine interaction mentioned here above, the nuclear
hamiltonian of gallium in SCGO also bears an additional quadrupole
interaction due to the electric field gradient (EFG) on the gallium sites. Following the
usual notations, the nuclear hamiltonian may be expressed as
\[
\mathcal{H}=-h \gamma \roarrow I\cdot (\tensor1+\tensor K)\cdot \roarrow %
H+\frac{h\nu _{Q}}{6}\left[ 3I_{z}^{2}-I^{2}+\eta (I_{x}^{2}-I_{y}^{2})%
\right]
\]
where $\roarrow H$ is the applied field, the principal axes of the magnetic
shift tensor $\tensor K$ are collinear with the direction of the nuclear
spin operators $I_{x}$, $I_{y}$, and $I_{z}$, $\nu _{Q}$ is the quadrupole
frequency and $0\leq \eta \leq 1$ is the quadrupole asymmetry parameter. The
main focus of the spectral analysis presented in this section is to complete
the study of Ref.~\onlinecite{AmitNMR} by determining the quadrupole
contribution to the nuclear hamiltonian of Ga($4f$), working in a frequency
range ($\nu _{rf}=40.454$ MHz) more appropriate for the observation of the
$^{69,71}$Ga($4f$) resonance lines. We study here the spectrum of the $p=0.95$
sample which yields sharper quadrupole features than the $p=0.90$ sample (or
of any of the other samples studied). The knowledge of the quadrupole parameters
reported in Tab.~\ref{tab:2} allows us, in the following section, to
separate the magnetic and the quadrupole contributions to the Ga($4f$)
spectrum and, hence, to safely evaluate the magnetic properties of the
kagom\'{e} bi-layer.

\begin{table}[b]
\caption{The quadrupole parameters of the $4e$ and $4f$ gallium nuclei in SCGO.
The $^{69}$Ga isotope yields a stronger quadrupole interaction compared to
the $^{71}$Ga isotope, since the quadrupole moments are in a ratio of $%
^{69}Q/^{71}Q=^{69}\protect\nu _{Q}/^{71}\protect\nu _{Q}=1.589$.}
\label{tab:2}%
\begin{ruledtabular}
\begin{tabular}{@{\extracolsep{0ptplus1fil}}cccc}
   & $^{71}\nu _{Q}$ (MHz) & $^{69}\nu _{Q}$ (MHz) & $\eta$   \\
\colrule
Ga($4e$)\footnotemark[1] & 20.5(3)               & 32.6(5)               & 0.050(35) \\
Ga($4f$)\footnotemark[2] &  2.9(2)               &  4.6(2)               & 0.005(6)  \\
\end{tabular}
\end{ruledtabular}
\footnotetext[1]{From Ref.~\onlinecite{AmitNMR}.}
\footnotetext[2]{This work.}
\end{table}

In the second part of this section we focus on Ga($sub$). We identify the
contribution of Ga($sub$) to the spectrum acquired at $\nu _{rf}=40.454$ MHz
and, on the basis of this contribution, provide a new method to evaluate the
Cr-concentration of the SCGO samples through Ga-NMR, especially well suited for
the very low dilutions.

\begin{figure}[t]
\includegraphics[width=8cm,bbllx=35,bblly=100,bburx=520,bbury=720]{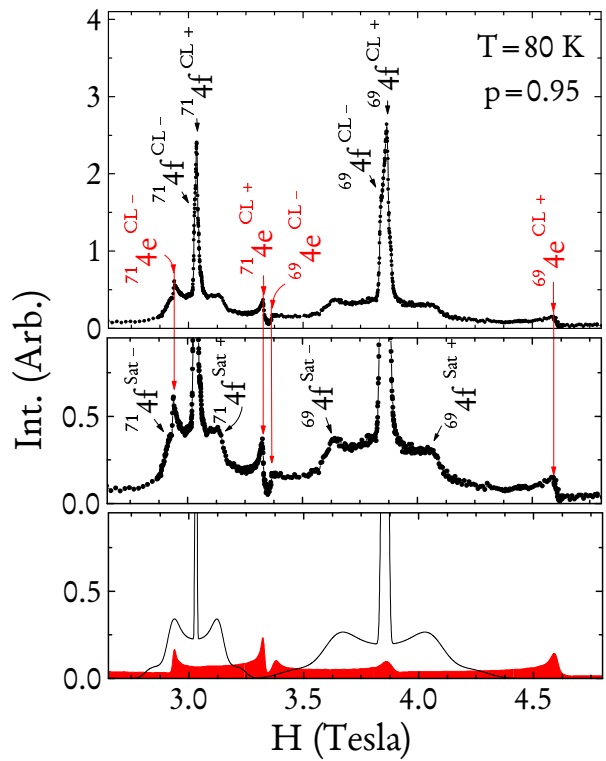}
\caption{\textit{Top panel}. A field sweep of $^{71}$Ga and $^{69}$Ga at $%
\protect\nu _{rf}=40.454$ MHz. The site and transition assignments of each
peak are described in the text. \textit{Middle panel}. The satellite transitions
in the $^{69,71}$Ga($4f$) line. The satellite singularities of the $4e$ line are
not resolved. \textit{Bottom panel}. A line simulation of $^{69,71}$Ga($4f$)
(solid line) and of $^{69,71}$Ga($4e$) (dark grey area) performed on the
basis of the quadrupole parameters of Tab.~\ref{tab:2}. The simulated
spectrum does not account for the Ga($sub$) contribution.}
\label{fig:5}
\end{figure}

\subsection{Spectral analysis of Ga($4f$)}

\label{Spectral analysis of Ga(4f)}

Neglecting the Ga-substituted sites, the Ga-NMR spectrum displays
four sets of lines corresponding to the two isotopes distributed on both
Ga($4e$) and Ga($4f$) sites. In a powder, for a given site and a given isotope,
the lineshape results from the distribution of the angles between the field and the
EFG principal axis. This yields singularities rather than well defined peaks
$-$ the so-called powder lineshape.

The quadrupole interaction of gallium nuclei in SCGO is a consequence of the coupling
of the nucleus to the EFG produced by the surrounding electronic charges,
here mainly the oxygen ions. As shown in Fig.~\ref{fig:2}, Ga($4e$) is
surrounded by a bi-pyramid of 5 oxygen ions. This environment has no local
cubic symmetry and results into a strong quadrupole frequency as was evidenced
in Ref.~\onlinecite{AmitNMR} (Tab.~\ref{tab:2}). In contrast with Ga($4e$), Ga($4f$) is
surrounded by a nearly ideal tetrahedron of oxygens ions, only slightly
elongated along the crystalline $\roarrow c$-axis. Since a regular
tetrahedron has a local cubic symmetry, we expect the quadrupole frequency
of Ga($4f$) to be $\nu _{Q}(4f)\ll \nu _{Q}(4e)$. This agrees with Ref.~\onlinecite{AmitNMR}
where the Ga($4f$) quadrupolar effects were not detected at a frequency of $\nu _{rf}=131$ MHz.
The asymmetry parameter $\eta $, that quantifies the deviation of the EFG from axial symmetry,
is in principle close to zero for both sites as they share a rotation axis along $\roarrow c$.

We present in the top panel of Fig.~\ref{fig:5} a field sweep spectrum, obtained at
$\nu _{rf}=40.454$ MHz and $T=80$ K for the $p=0.95$ sample. The sweep, from $2.65$
T to $4.80$ T, covers the spectrum of both isotopes and yields the expected
features. We first focus on the $^{69,71}$Ga($4e$) line. A powder sample simulation of the
$4e$-line is presented in the bottom panel of Fig.~\ref{fig:5} to emphasize its contribution
to the spectrum. Each $4e$-isotope, i.e. $^{71}$Ga($4e$) and $^{69}$Ga($4e$), exhibits
a resonance in a wide range of fields, with multiple quadrupole peaks and steps associated
to the three nuclear Zeeman transitions $3/2\longleftrightarrow 1/2$, $1/2\longleftrightarrow -1/2$ and
$-1/2\longleftrightarrow -3/2$. As evidenced in Fig.~\ref{fig:5}, in this field window only the $1/2\longleftrightarrow -1/2$
resonance transition is resolved yielding for each $4e$-isotope two singularities known as the central line singularities
(labeled respectively by $CL-$ and $CL+$ in Fig.~\ref{fig:5}). These two singularities are the boundaries of the
central line splitting of width $\approx \nu _{Q}^{2}/2\gamma
\nu _{rf}$.\cite{Baugher} Compared to the working frequency of $\nu _{rf}=131$ MHz
used in Ref.~\onlinecite{AmitNMR}, the much lower frequency of $\nu _{rf}=40.454$ MHz
allows here to spread the $4e$ line over a wider field-range.
The $^{69,71}$Ga($4e$) contribution to the spectrum is
a nearly constant background of small amplitude compared to the $^{69}$Ga($4f$)
and $^{71}$Ga($4f$) lines $-$ the two prominent peaks of Fig.~\ref{fig:5} $-$
hence there is a better contrast between the sites. Furthermore, overlap
between the Ga($4e$) and the Ga($4f$) lines is also minimized at this radio-frequency.

\begin{figure}[b]
\includegraphics[width=4.5cm,bbllx=50,bblly=200,bburx=490,bbury=730,clip=]{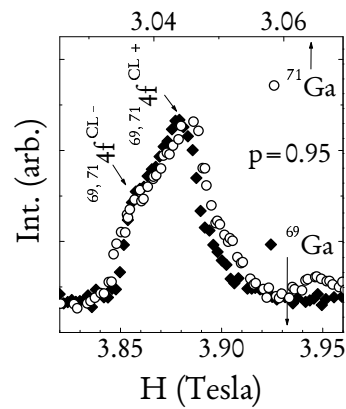}
\caption{Field sweeps of the $^{69}$Ga($4f$) and the $^{71}$Ga($4f$) lines
at $T=150$ K. The closed (open) symbols represent $^{69}$Ga ($^{71}$Ga). The
top abscissa is scaled to the bottom by a factor $^{69}(Q^{2}/\protect%
\gamma )/^{71}(Q^{2}/\protect\gamma )=3.209$ to evidence the quadrupole
structure in the lines. The hump at $\approx 3.065$ T is $^{71}$Ga($sub$).}
\label{fig:6}
\end{figure}

We now focus on the $^{69,71}$Ga($4f$) line of Fig.~\ref{fig:5}. In these
favorable experimental conditions, it can be seen that both $4f$ peaks are
flanked by two shoulders, the so-called satellite singularities
(labeled respectively by $Sat-$ and $Sat+$ in the middle panel of Fig.~\ref{fig:5}),
associated with the $3/2\longleftrightarrow 1/2$ ($Sat-$) and the $-1/2%
\longleftrightarrow -3/2$ ($Sat+$) nuclear transitions. The detection of the
satellite singularities is a clear indication of the existence of quadrupole
effects in the $^{69,71}$Ga($4f$) spectrum, implying also a central line
splitting for Ga($4f$). To emphasize this central line splitting, we present in Fig.~\ref{fig:6}
narrow field sweeps of both isotopes at $T=150$ K. The sweep ranging from
$3.820$ T to $3.960$ T covers the $^{69}$Ga($4f$) line (bottom abscissa) whereas
the sweep ranging from $3.026$ T to $3.069$ T covers the $^{71}$Ga($4f$) line
(top abscissa). As shown, the two central lines match since we expanded the field window
of $^{71}$Ga($4f$) by a factor $^{69}(Q^{2}/\gamma )/^{71}(Q^{2}/\gamma )=3.209$, i.e. by the ratio of the
central line quadrupolar widths expected for the two isotopes. Fig.~\ref{fig:6} also indicates
that the structure observed in the $4f$ line is dominated,
at these temperatures, by quadrupole effects. In particular, the anisotropy in the shift $\tensor K$,
which could in principle yield additional structure to the line, is small,
i.e. the shift is isotropic ($\tensor K\equiv K$). This conclusion is
further supported by a calculation, presented below, of \textit{all} the field singularities.

\begin{table}[t]
\caption{The field positions of the $4f$ quadrupole singularities in Fig.~\ref
{fig:5}. $H_{sing}$ are the fields where each singularity was found
experimentally and $H_{sing}^{calc}$ the calculated fields where the
singularities are expected. The missing $H_{sing}^{calc}$ correspond to the field singularities
used to perform the calculation.}
\label{tab:3}%
\begin{ruledtabular}
\begin{tabular}{@{\extracolsep{0ptplus1fil}}ccccc}
Sing.       & $^{71}H_{sing}$ (T) & $^{69}H_{sing}$ (T) &  $^{71}H_{sing}^{calc}$ (T) & $^{69}H_{sing}^{calc}$ (T) \\
\colrule
$4f^{CL-}$  & 3.025(2)            & 3.838(1)            & 3.026(2)                    & 3.838(2)                   \\
$4f^{CL+}$  & 3.036(1)            & 3.865(1)            & 3.035(2)                    & $-$                        \\
$4f^{Sat-}$ & 2.924(9)            & 3.628(8)            & 2.921(9)                    & $-$                        \\
$4f^{Sat+}$ & 3.137(7)            & 4.066(8)            & 3.138(7)                    & $-$
\end{tabular}
\end{ruledtabular}
\end{table}

We now turn to the evaluation of the quadrupole parameters of $^{69,71}$Ga($4f$).
This was carried out using the field positions $H_{sing}$ of the
quadrupole singularities determined from Fig.~\ref{fig:5} and reported in
Tab.~\ref{tab:3} which depend only on the three parameters $\nu _{Q}(4f)$,
$\eta (4f)$ and $K$, provided the shift is isotropic: $H_{sing}=H(\nu_{Q},\eta ,K)$. \cite{Baugher}
We evaluated the $4f$ quadrupole parameters
from the singularity positions of $^{69}$Ga($4f$)$^{Sat-}$, $^{69}$Ga($4f$)$
^{CL+}$ and $^{69}$Ga($4f$)$^{Sat+}$. At $T=80$ K, the shift is $K=0.0285(5)$.
The extracted quadrupole parameters are presented in Tab.~\ref{tab:2} and
are found in agreement with expectations. To confirm the isotropic nature of
the shift, we performed a self-consistency test by calculating the field
positions of all the quadrupole singularities in Fig.~\ref{fig:5}. This is
readily done using the three parameters $\nu _{Q}(4f)$, $\eta (4f)$ and $K$
determined previously and the set of equations of Ref.~\onlinecite{Baugher}.
As shown in Tab.~\ref{tab:3}, the calculated positions are in excellent
agreement with the experimental ones.

Finally, concerning the quadrupole interaction in all the other SCGO samples
studied, no appreciable change was evidenced in $\nu _{Q}(4f)$ and $\eta(4f)$.
In conclusion, the Ga($4f$) central line is narrow enough in all the samples to allow to
follow accurately the $T$-variation of the shift and of the linewidth.

\subsection{Gallium substituted on the Cr sites}

\begin{figure}[b]
\includegraphics[width=7cm,bbllx=20,bblly=270,bburx=550,bbury=710,clip=]{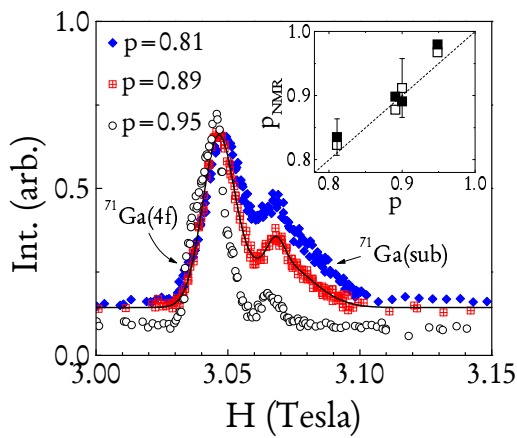}
\caption{A field sweep centered on the $^{71}$Ga($4f$) central line performed
at $T=150$ K. The solid line is a multiple gaussian fit. \textit{Inset}. $%
p_{NMR}$ versus $p$. The open and closed symbols are respectively the
Cr-concentration evaluated by direct integration and by multiple gaussian
fits. The dashed line is the expected value from neutron/X-ray diffraction.}
\label{fig:7}
\end{figure}

In Fig.~\ref{fig:7} we compare the $^{71}$Ga spectra obtained for the $p=0.81$,
$0.89$, $0.95$ samples at $T=150$ K and $\nu _{rf}=40.454$ MHz. Care was
taken to ensure the same experimental conditions. The field sweeps of
Fig.~\ref{fig:7} cover a narrow field window $\delta H$ (from $3.00$ T to $3.15$
T) centered on the central line of $^{71}$Ga($4f$) ($H\approx 3.05$T). Two
features differ between the samples. 1) The line at $H\approx 3.065$T whose
integrated area decreases with increasing $p$, which is therefore $^{71}$Ga($%
sub$). 2) The constant background $b(p)$ which decreases with increasing
$p$. $^{71}$Ga($sub$) and part of $b(p)$ are therefore all the contribution of
the non-stoichiometric gallium to the $^{71}$Ga spectrum.\cite{69Ga}

Since the non-stoichiometric gallium is present on the three Cr sites of SCGO,
one would expect to detect three gallium lines and not a unique Ga($sub$)
line. To gain insight into this problem, we studied the spectrum of the
parent and non magnetic compound SrGa$_{12}$O$_{19}$ ($p=0$). The spectrum
allows to evidence sizeable quadrupole effects on the $12k$ ($^{71}\nu
_{Q}=5.1(1)$ MHz and $\eta =0.47(3)$) and on the $4f_{vi}$ ($^{71}\nu _{Q}=1.6(1)$
MHz and $\eta =0.03(5)$) substituted sites, otherwise non measurable in the
magnetic compounds ($p\neq 0$).\cite{TheseLaurent} The quadrupole effects on
the $2a$ site are negligible ($^{71}\nu _{Q}\approx 0.3$ MHz and $\eta\approx 0$).
Turning back to the spectra of Fig.~\ref{fig:7}, the $^{71}$Ga($sub$) line
is the unresolved sum of the $2a$ line and of the $12k$ and $4f_{vi}$ quadrupole
central lines of substituted gallium, whereas the background $b(p)$, in
addition to the Ga($4e$) and Ga($4f$) satellites contributions mentioned in
Sec.~\ref{Spectral analysis of Ga(4f)}, contains the sum of the satellite
contributions of the substituted gallium.

In order to evaluate the Cr-concentration of our samples using Ga-NMR
($p_{NMR}$), we determine the intensities $I_{4f}$ and $I_{sub}(p)$ of respectively
the $^{71}$Ga($4f$) central line and the $^{71}$Ga($sub$) line of Fig.~\ref{fig:7}.
To do so, we evaluate the integrated area of both lines corrected
for the transverse exponential relaxation ($T_{2}(sub)=78(2)\mu $s and $%
T_{2}(4f)=50(1)\mu $s). $I_{sub}(p)$ and $I_{4f}$ are respectively
proportional to the amount of gallium present on the three substitution
sites and on the $4f$ site. The lines are well separated in the $p=0.95$
sample, so that $I_{sub}(p)$ and $I_{4f}$ are evaluated separately. For the more
diluted samples, the intensity $I_{4f}$ is fixed to the $p=0.95$ value as the amount of gallium on the $4f$ site
is independent on Cr-concentration. $I_{sub}(p)$ was then evaluated either
by direct integration of the spectrum over the field window $\delta H$ and
subsequent subtraction of both $b(p)\delta H$ and $I_{4f}$, or by reproducing
all the lines with a minimal $3$ gaussian fit and an additional constant
background $b(p)$ (one gaussian for the $^{71}$Ga($4f$) line with an
intensity fixed to the $p=0.95$ value, and two gaussians for $^{71}$Ga($sub$),
the area of which yields $I_{sub}$). From the ratio $I_{sub}/I_{4f}$, we
can extract $p_{NMR}$. In the inset of Fig.~\ref{fig:7} we show the
variation of $p_{NMR}$ with the concentration $p$ determined by X-ray and
neutron diffraction. Whatever the method employed (integration or multiple
gaussian fits), we notice a perfect agreement between the different
characterization methods. Since $p_{NMR}$ is equal to $p$, we conclude that
the actual chemical content corresponds to the nominal concentration.

Incidentally, Ga-NMR proves to be a quite accurate tool for determining the
Cr-concentration in SCGO, especially for dilutions as low as $5\%$ or less
where the other techniques are beyond their sensitivity limit. For the samples with $p>0.95$, we
did not find any further decrease of the $I_{sub}$ value which confirms
\textit{locally} that it is impossible to synthesize samples beyond the $p=0.95$
limit found through X-ray diffraction.

\section{Dilution and $T$-dependence of the Ga($4f$) spectrum}

\label{Dilution and $T$-dependence of the Ga($4f$) spectrum}

To appreciate the experimental evidence of the properties we discuss in the
next sections, we present here the dilution and the $T$-dependence of the raw
$^{71}$Ga($4f$)-NMR spectra. The main relevant physical parameters of
interest are the shift of the line, which as we establish allows us to
probe directly the susceptibility of the kagom\'{e} bi-layer, and the
linewidth which in contrast yields information on the distribution of the
internal local fields in the kagom\'{e} bi-layer. We also briefly describe
the analysis performed to extract these spectral parameters.

\subsection{Experimental data}
\label{exp data}

\begin{figure}[t]
\includegraphics[width=7cm,bbllx=40,bblly=260,bburx=540,bbury=705,clip=]{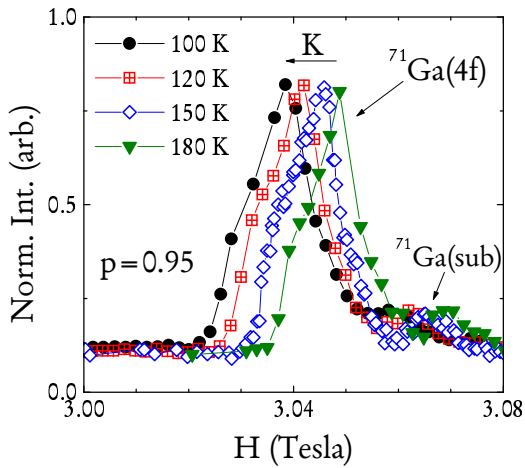}
\caption{Typical high-$T$ field sweeps of the $^{71}$Ga($4f$) central line.}
\label{fig:8}
\end{figure}

The Ga($4f$) spectra were recorded in a wide $T$-range, from $5$ K to $410$
K. Up to $200$ K, the spectra were obtained in a sweep field set up with a
working frequency of $\nu _{rf}=40.454$ MHz. For $150$ K$\leq T\leq 410$ K,
the spectra were recorded in a static field of $\approx 7$ T by a
swept-frequency variant of the field step spectroscopy method.\cite{Clark}
Our data can be sorted in two distinct temperature domains, typically $T\geq
100$ K and $T<100$ K.

A typical series of high-$T$ spectra ($100$ K$\leq T\leq 180$ K) is
presented in Fig.~\ref{fig:8} for the $p=0.95$ sample. The field window (from
$3.00$ T to $3.08$ T) covers the central line of $^{71}$Ga($4f$). Fig.~\ref
{fig:8} is representative of the evolution of Ga($4f$) at $T\geq 100$ K for
all the samples studied. The line shape is nearly constant with temperature, as
dominated by $T$-independent quadrupole effects, and shifts towards the low
fields with decreasing temperature. The shift $K$, which is experimentally
measured by the position of the $^{71}$Ga($4f$)$^{CL+}$ singularity and
performing minor quadrupole corrections (see Sec. \ref{Ga-NMR spectrum}),
therefore increases when the temperature is lowered.

\begin{figure}[b]
\includegraphics[width=7cm,bbllx=30,bblly=185,bburx=425,bbury=715,clip=]{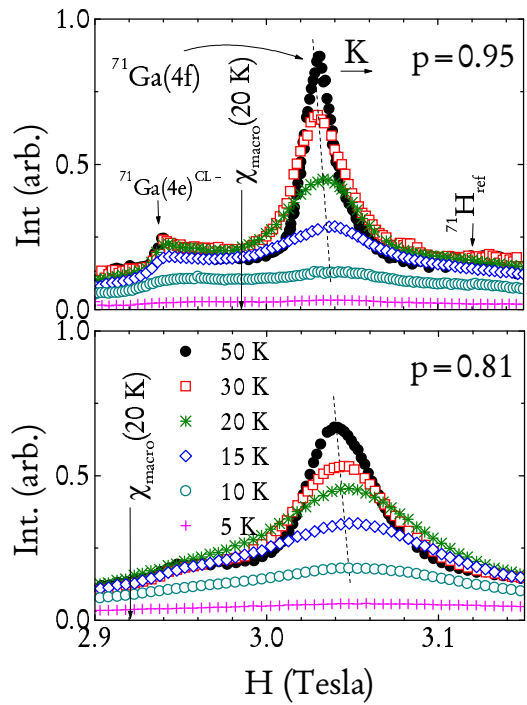}
\caption{Low-$T$ field sweeps of $^{71}$Ga($4f$) for the $p$=0.95 sample (top
panel) and the $p$=0.81 sample (bottom panel). The dashed line is a guide to
the eye.}
\label{fig:9}
\end{figure}

The $T$-dependence of both $K$ and the width changes at low-$T$. We present
in Fig.~\ref{fig:9} spectra ranging from $5$ K to $50$ K for the $p=0.95$
sample (top) and the $p=0.81$ sample (bottom). The field window covers the
$^{71}$Ga($4f$) line, but is larger than in Fig.~\ref{fig:8}, from $2.90$ T
to $3.15$ T. Three important features should be noticed.

\begin{figure}[t]
\includegraphics[width=7cm,bbllx=40,bblly=240,bburx=520,bbury=695,clip=]{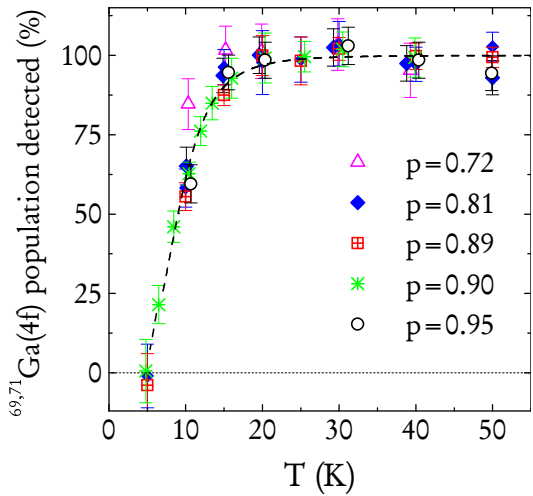}
\caption{The  $^{69,71}$Ga($4f$) population detected by NMR. For each sample, the
integrated area of the $4f$ line is normalized by the integrated area measured in
the $20-50$ K temperature range, where it is found constant in temperature.
The Ga($4e$) contribution, constant in temperature, was subtracted. $T_{2}$ corrections
were found quite small and do not affect estimates below 50 K.}
\label{fig:10}
\end{figure}

\begin{enumerate}
\item  The line now shifts toward the high fields and hence $K$ is
decreasing with the temperature. However $K$ does not reach the zero value,
i.e. the $^{71}$Ga($4f$) line is never centered at the reference value
$^{71}H_{ref}=\nu _{rf}/^{71}\gamma =3.116$ T. The comparison of the two
series of spectra establishes that the change in the shift direction is not
affected by the dilution.

\item  The width increases with decreasing temperature. The quadrupole
structure that can still be noticed on the $T=50$ K spectrum of the $p=0.95$
sample, is progressively washed away as temperature is lowered. At $T=20$ K,
the line is marked by a smooth symmetric broadening of gaussian nature. This
broadening is more important for the $p=0.81$ sample than for the $p=0.95$
sample, i.e. the width is sensitive to the dilution. The $^{71}$Ga($sub$)
line is no longer resolved in this $T$-domain.

\item  A last feature we would like to underline is the abrupt decrease of
the detected $^{69,71}$Ga($4f$) nuclear population at $T<15$ K in both samples.
The nuclear population, proportional to the integrated area of the $4f$
line, actually decreases in the same way in all the samples studied (Fig.~\ref
{fig:10}). It therefore originates from a dilution-independent mechanism.
The present result confirms the conclusions of the previous Ga($4f$)-NMR
study on the $p=0.90$ sample.\cite{Philippe} There, the wipe-out of the
intensity was assigned to originate from the intrinsic high dynamics of the
kagom\'{e} bi-layer spin system when $T\rightarrow 0$ K. This effect,
explicitly evidenced in SCGO by $\mu $SR measurements,\cite{Uemura,AmitMuSR}
is taken to be the signature of a spin liquid-like ground state. Because of this
abrupt decrease of the NMR signal,  $^{69,71}$Ga($4f$) is not a suitable probe for
$T<10-15$ K.
\end{enumerate}

The comparison of the spectra of all the samples enables us in the next
sections to easily separate the properties related to the dilution from the
intrinsic properties independent on the dilution.

\subsection{NMR line analysis}
\label{line analysis}

Compared to Ref.~\onlinecite{Philippe}, we employed here a different method to
evaluate the shift $K$ and the width $\Delta H$ of the spectra of Fig.~\ref
{fig:9}. (i) For all the samples, we choose a $^{69,71}$Ga($4f$) high-$T$
reference spectrum dominated by $T$-independent quadrupole effects
($\mathcal{S}_{Q}(H)$) (ii) In order to recover the low-$T$ broadening of
gaussian nature, $\mathcal{S}_{Q}(H)$ is convoluted by a normalized gaussian
function $G(H)$ of width $\Delta H$. Since it is impossible to fit
separately the substituted and non-substituted sites, we explicitly assume
that for the highly diluted samples both sites broaden identically. This is
likely the case in view of the hyperfine coupling paths. The low-$T$
spectrum, $\mathcal{S}_{M}(H)$, is therefore simply reproduced by
\[
\mathcal{S}_{M}(H)=a\left[ y_{0}+\int_{-\infty }^{+\infty }G(H-H^{\prime })%
\mathcal{S}_{Q}(H^{\prime }+H_{s})dH^{\prime }\right]
\]
where $H_{s}$ is the field-shift of $\mathcal{S}_{Q}(H)$, from which we
extract $K$. Since the $^{69,71}$Ga($4f$) line lies on a constant background
that results mainly from the contribution of the second gallium site Ga($4e$) (see
Fig.~\ref{fig:5}), the amplitude $a$ and the constant background $y_{0}$ are
employed to readjust the relative spectral weights of the $4e$ and of the
$4f$ lines when the transverse relaxation $T_{2}$ corrections varies
differently on the two gallium sites.\cite{T2} We also restricted our fits
to the left side of the line, very sharp at high-$T$ and which has the
major advantage of not being affected by the Ga/Cr substitutions. Concerning the
values of $K$, we did not find any significant difference with the method
employed in Ref.~\onlinecite{Philippe}.

\subsection{Magnetic contribution to the Ga($4f$)-NMR line: shift and width}
\label{Magnetic contribution to the NMR line}

We briefly recall here the relationship between the contribution of each gallium
nucleus to the Ga($4f$)-NMR line and its local magnetic environment in
order to underline what can be exactly probed through Ga($4f$)-NMR (more details
can be found in Ref.~\onlinecite{AmitNMR,TheseLaurent,LaurentHFM2000}).

Each Ga$(4f)$ nucleus is coupled to its Cr$(12k)$ and Cr$(2a)$ nearest neighbors (n.n.)
through a Ga-O-Cr hyperfine interaction with an hyperfine constant
$\mathcal{A}$.\cite{hyperfine} We suppose that the susceptibility in the kagom\'e bi-layer
varies from Cr site to Cr site and label it, in a generic manner, by $\chi$.
A Ga$(4f)$ at site $i$ will contribute to the NMR spectrum
at a position depending upon the number of the n.n. occupied Cr sites and their
susceptibility $\chi$. This corresponds
to the shift $K^{(i)}$ in the NMR spectrum for a gallium at site $i$
\[
K^{(i)}=\sum_{\text{occupied n.n.Cr}(12k,2a)}\mathcal{A}\chi
\]
(the chemical shift is negligible). The average shift of the NMR line, that we label by $K$
in the following sections, is simply related to the average susceptibility $\overline{\chi}$ over all
the Cr sites. As we establish, the shift corresponds to the frustrated susceptibility
$\chi _{frustr}$, so that $K\propto \chi_{frustr}$. Instead, the spatial distribution of
$K^{(i)}$ around $K$ defines the magnetic width of the Ga-NMR spectrum and reflects the existence of a
spatial distribution of $\chi$ (as we detail in Sec.~\ref{Dilution effects through the Ga(4f)-NMR width},
a spatial distribution of the hyperfine constant or a distribution
related to the presence of Cr-vacancies in the nuclear environnement is
negligible and cannot justify the low-$T$ broadening observed).

\section{Kagom\'e bi-layer susceptibility: evidence for correlations effects}
\label{section:kagome bi-layer susceptibility}

As mentioned in the previous section, the shift $K$ measures the average
kagom\'{e} bi-layer susceptibility that proves to be unobservable at low-$T$
through the macroscopic susceptibility measurements. We show here that
the $T$-dependence of $K$ gives a sharp evidence for a maximum in the
susceptibility $\chi _{frustr}$ around $50$ K, whatever the amount of dilution.
The comparison of our results on the SCGO compound with the different models
existing in the literature, does not allow us to totally validate or invalidate
them. Our observations rather favor an image, supported by recent neutron
diffraction data, where the short range magnetic correlations play a central role.

\subsection{An intermediate temperature scale}

\begin{figure}[t]
\includegraphics[width=6.5cm,bbllx=40,bblly=85,bburx=500,bbury=705,clip=]{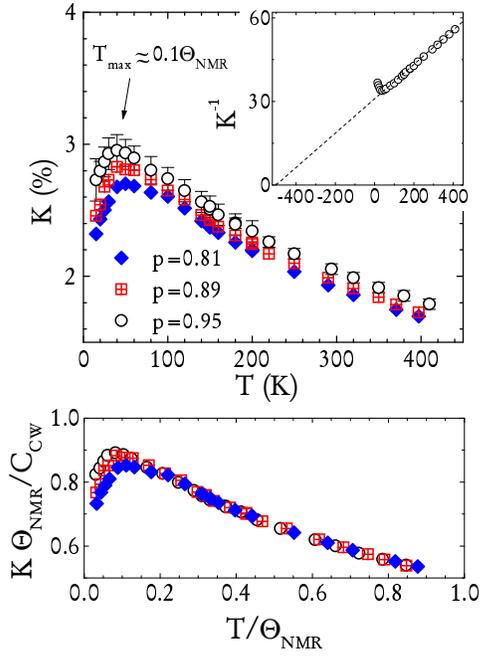}
\caption{\textit{Top panel}. $K$ versus $T$ down to 15 K. A minor second-order quadrupole correction has been
performed. \textit{Inset}. $K^{-1}$ versus $T$ for $p=0.95$. \textit{Bottom panel}. $K$ versus $T$ plotted using
reduced units. $C_{CW}$ and $\Theta _{NMR}$ are respectively the Curie-Weiss constant and temperature extracted
from the high-$T$ Curie-Weiss fit to $K^{-1}$ described in the text.}
\label{fig:11}
\end{figure}

The $T$-dependence of $K$ for the $p=0.81,0.89,0.95$ samples is presented in
Fig.~\ref{fig:11}. The figure is the quantitative evaluation of the line
shift of Fig.~\ref{fig:8} and Fig.~\ref{fig:9}. $K$ increases following a
Curie-Weiss law (see below) up to a temperature of $T_{max}=40-50$ K, where
it reaches a maximum. Below $T_{max}$, $K$ decreases, but
does not reach $0$ for the observed temperatures ($T\geq 15$ K). The
decrease in $K$ is slightly more pronounced with increasing dilution. For
the $p=0.81$ sample $K(T=15$ K$)/K(T_{max})\approx 0.86$, instead of $0.92$ for
the $p=0.95$ sample. Although for $T<15$ K the measurement of $K$ is less
significative since a wipe out of the intensity occurs, we observed in the
$p=0.90$ sample only a decrease of $K(T=7$ K$)/K(T_{max})\approx 0.77$.\cite{Philippe}

The typical $T$-variation of $K^{-1}$ is presented in the inset of Fig.~\ref
{fig:11} for the $p=0.95$ sample. The high-$T$ variation of $K^{-1}$ ($T\geq
100$ K) is linear for all the samples, suggesting a Curie-Weiss behavior. A
linear extrapolation to $K^{-1}=0$ yields the Curie-Weiss temperature as
determined by NMR ($\Theta _{NMR}$). The values of $\Theta _{NMR}$ are
$453(30)$ K, $469(27)$ K and $484(25)$ K for the $p=0.81,0.89,0.95$ sample
respectively. We find here in $K$ a well known property of the
susceptibility of the geometrically frustrated compounds: the Curie-Weiss behavior
continues to subsist at temperatures $T\ll \Theta _{NMR}$.

Assuming the Curie-Weiss value of the kagom\'{e} bi-layer to be $\left\langle%
{z}\right\rangle J_{bi-layer}S(S+1)/3k_{B}$, where $\left\langle{z}%
\right\rangle =5.14$ is the average number of nearest neighbors for a
Cr site of the kagom\'e bi-layer as the Cr-environments of the Cr($2a$) and the Cr(%
$12k$) sites differ, we find an average exchange constant of $J_{bi-layer}\approx
80$ K. As shown in Fig.~\ref{fig:4}, this value is close to the exchange
constant $J=76(3)$ K observed in Cr$_{2}$O$_{3}$ where the Cr$^{3+}$ ions have a
local octahedral environment similar to Cr($12k$).\cite{Samuelsen} We also
gained more insight on the couplings of the kagom\'e bi-layer through a
mean-field analysis of the high-$T$ shift, detailed in Ref.~%
\onlinecite{LaurentHFM2000}, that allowed us to evaluate the Cr($12k$)-Cr($%
2a$) and the Cr($12k$)-Cr($12k$) couplings (their values are reported on
Fig.~\ref{fig:4}). Turning back to the NMR Curie-Weiss temperature, $\Theta _{NMR}$ is of the same order
as the macroscopic Curie-Weiss temperature, although
smaller with an increasing difference at higher Cr-concentrations. This
difference is related to the fact that the susceptibility of the Cr($4f_{vi}$%
)-Cr($4f_{vi}$) isolated spin pairs also contributes to $\chi _{macro}$, but
is absent in $K$ since the Ga($4f$) nuclei probe only the susceptibility of
the kagom\'{e} bi-layer (Sec.~\ref{Evidence for a two component macroscopic
susceptibility}).

The linear behavior of $K$ for $T\ll \Theta _{NMR}$ and the deviation from
this behavior that results in a maximum in $K$ versus temperature, is a common feature
of the three samples. It then stems from a physics robust on dilution,
therefore related to an intrinsic property of the kagom\'{e} bi-layer.
$T_{max}$ is a new temperature scale for SCGO since it differs substantially
from the two known characteristic temperatures, the freezing temperature
$T_{g}$ ($T_{max}/T_{g}\approx 10$) and the Curie-Weiss temperature
($T_{max}/\Theta_{NMR}\approx 0.1$).

\subsection{Evidence for a two component macroscopic susceptibility}

\label{Evidence for a two component macroscopic susceptibility}

\begin{figure}[t]
\includegraphics[width=7cm,bbllx=75,bblly=330,bburx=530,bbury=705,clip=]{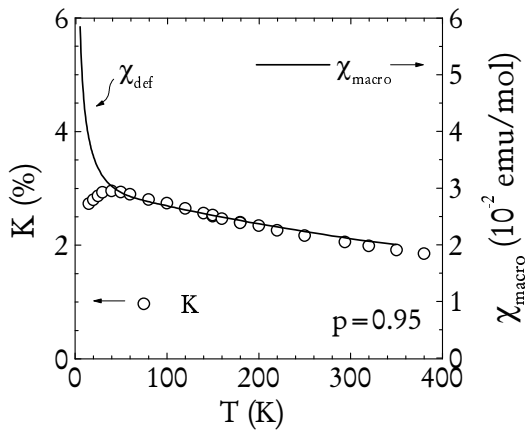}
\caption{$K$ (\textit{left}) and $\chi _{macro}$ (\textit{right})
for the p=0.95 sample. $K$ and $\chi _{macro}$ do not follow exactly
the same law at high-$T$, since $\chi _{macro}$ probes also the
susceptibility of the Cr($4f_{vi}$)-Cr($4f_{vi}$) spin pairs.}
\label{fig:12}
\end{figure}

We compare in Fig.~\ref{fig:12} the $T$-dependence of $K$ (left y-axis) and
of $\chi _{macro}$ (right y-axis) for the $p=0.95$ sample. The discrepancy
between the low-$T$ behavior of $K$ and $\chi _{macro}$ is quite clear since
instead of a maximum in the susceptibility, $\chi _{macro}$ exhibits a Curie-like
law at low-$T$. The difference between $K$ and $\chi _{macro}$ is also
perceptible directly on the spectra of Fig.~\ref{fig:9} where an arrow
indicates the approximate position where the maximum of the $^{71}$Ga($4f$) line
should occur if $K\propto \chi _{macro}$. Instead of a model where $\chi
_{macro}$ has a single component, our results establish that $\chi _{macro}$
yields at least two distinct contributions. The first one is $\chi _{frustr}$,
spatially uniform over the kagom\'{e} bi-layer and that is
reflected in $K$ (Sec.~\ref{Magnetic contribution to the NMR line}).
A second one, non-uniform over the Cr sites, that is necessary to explain the Curie-like
upturn observed at low-$T$. We label this contribution by $\chi _{def}$,
since in Sec.~\ref{section:dilution effects} we establish that $\chi _{def}$
is the susceptibility of the magnetic defects generated by the dilution of the kagom\'e bi-layer.
The comparison of Fig.~\ref{fig:12} establishes on an experimental ground the conjecture
by Schiffer \textit{et al.} of a two component macroscopic susceptibility.\cite{Schiffer}

The macroscopic susceptibility probes both contributions at low-$T$. If we
also take into account the susceptibility of the Cr($4f_{vi}$)-Cr($4f_{vi}$)
isolated spin pairs ($\chi _{pair}$), the three contributions to $\chi _{macro}$ are
\begin{equation}
\chi _{macro}=\chi _{frustr}+\chi _{pair}+\chi _{def}  \label{eqn:sus}
\end{equation}
The dominant Curie susceptibility of the defects $\chi _{def}$ at low-$T$
prevents to probe $\chi _{frustr}$ through macroscopic
measurements.

Using Eq.(\ref{eqn:sus}), we can fit the $T$-dependence of $\chi _{macro}$
given some simple remarks and minor assumptions. The $T$-dependence
of $\chi _{frustr}$ is known through $K$, and for this reason the fit was performed in
the $15$ K$\leq T\leq 350$ K range, where we have a full intensity in the NMR signal.
This $T$-range of the fit leads to an inaccuracy in the determination of a possible low-$T$ Curie-Weiss temperature
of $\chi _{macro}$, hence we assume a pure Curie component $\chi _{def}=C_{def}/T$ to fit the
low-$T$ upturn. The susceptibility $\chi _{pair}$ is quantitatively derived
analytically (see Appendix). The only unknown parameters of Eq.(\ref{eqn:sus})
are the effective moment $\mu _{eff}(4f_{vi})$ of the Cr$^{3+}$ ions of $\chi _{pair}$,
the Curie constant $C_{def}$ of the defects and the hyperfine constant $\mathcal{A}$.

\begin{figure}[b]
\includegraphics[width=7cm,bbllx=40,bblly=250,bburx=520,bbury=705,clip=]{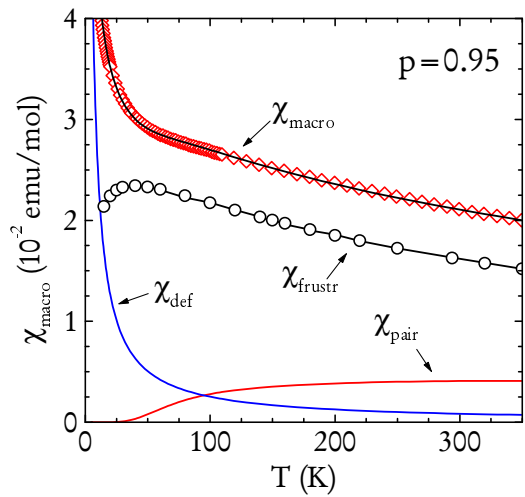}
\caption{The three contributions to $\chi _{macro}$ obtained through the fit
described in the text: the kagom\'e
bi-layer susceptibility ($\protect\chi _{frustr}$), the Cr($4f_{vi}$)-Cr($%
4f_{vi}$) spin pairs susceptibility ($\protect\chi _{pair}$) and the Curie
susceptibility induced by dilution ($\protect\chi _{def}=C_{def}/T$).}
\label{fig:13}
\end{figure}

The fit to $\chi _{macro}$ for the $p=0.95$ sample is presented in Fig.~\ref
{fig:13} and reproduces correctly the evolution of the experimental
macroscopic susceptibility in the $T>15$ K range. In Tab.~\ref{tab:6} we
have reported the fitting parameters extracted from the three samples
$p=0.81,0.89,0.95$. The effective moment is close to the $3.87\mu _{B}$ value
expected for a Cr$^{3+}$ ion, and the Curie contribution $C_{def}$ is found
to decrease with increasing $p$. The values of $C_{def}$ presented here,
even if extracted from a fit which does not cover the $5$ K$\leq T\leq 15$ K
range where $\chi _{def}$ dominates, nevertheless are  in good
agreement with those of Tab.~\ref{tab:4} where the $C_{def}$ value is
dominantly determined by the very low-$T$ region fit of $\chi _{macro}$
(Sec.~\ref{section:dilution effects}). Finally, $\mathcal{A}$ is nearly constant
with $p$, confirming the assumption of Sec.~\ref{section:samples} that
the site dilution does not alter in a significant way the couplings in SCGO, hence
that the Ga-O-Cr hyperfine interaction is also constant with the dilution.

\begin{table}[t]
\caption{The fitting parameters $\mu _{eff}(4f_{vi})$, $C_{def}$ and $\mathcal{A}$ for the $
p=0.81,0.89,0.95$ samples used to reproduce $\chi _{macro}$ by Eq.(\ref{eqn:sus}).}
\begin{ruledtabular}
\begin{tabular}{@{\extracolsep{0ptplus1fil}}cccc}
$p$ & $ \mu _{eff}(4f_{vi}) (\mu _{B})$ & $C_{def}$ (emu K/mol) & $\mathcal{A}\times 10^{3}$ (Oe$/\mu _{B}$) \\
\colrule
$0.81$ & $4.0(6)$ & $0.51(12)$ & 3.4(4)\\
$0.89$ & $4.5(4)$ & $0.45(10)$ & 3.8(4)\\
$0.95$ & $4.3(2)$ & $0.25(5)$  & 3.9(2)\\
\end{tabular}
\end{ruledtabular}
\label{tab:6}
\end{table}

\subsection{Discussion}

We compare here our experimental data for the intrinsic susceptibility $\chi _{frustr}$ of the
frustrated lattice reflected in $K$ to the calculations
performed for the kagom\'{e} and the pyrochlore networks through different
approaches.

Within a classical frame where the minimization of the Heisenberg
hamiltonian is constructed on the basis of non magnetic triangles and
tetrahedras, different authors have simulated the susceptibility of the
kagom\'e and the pyrochlore network by Monte-Carlo calculations.\cite
{MonteCarlo,Schiffer,Moessner} All these models suggest a Curie-Weiss
susceptibility which extends to low-$T$ ($T\ll\Theta$), in agreement with
the high-$T$ behavior of $K$ ($T>100$ K). Although these models do not
predict a maximum in the susceptibility versus temperature, all agree with our results on the absence
of a Curie upturn at low-$T$ as is detected in $\chi _{macro}$.

The susceptibility for the $S=1/2$ kagom\'{e} network has been calculated by an
exact diagonalization of the Heisenberg hamiltonian.\cite{Lhuillier} As
mentioned in the introduction, the ground state is here constructed on
non magnetic spin singlets and a ``gap'' ($\Delta $) is predicted between
the singlet ground state and the magnetic spin triplet state. The calculated
susceptibility hence varies as $\sim exp(-\Delta /T)$ at low-$T$, with a
maximum in the susceptibility occurring at $\Delta \approx 0.1J$. However, it seems difficult
to reconcile this maximum with our results. First, we would expect to
observe in $K$ a much sharper decrease in temperature than what is observed
experimentally. Second, the energy scale of $\Delta $ itself does not agree
with our data. In fact, taking $J\approx 80$ K, the maximum of $K$ falls at
$T_{max}\approx 0.5J>\Delta $ (note that $\Delta $ should be weaker than
$0.1J $ for $S=3/2$ spins or even disappear).\cite{Lhuillierprivate} The
interpretation of the observed maximum for SCGO has to be researched
elsewhere than in the gap predicted by this model. We stress that the Ga(4f)-NMR
intensity loss does not allow us to probe deeply the
predictions of this model which places the gap in the $T<15$ K range. But,
clearly, our data reveal that other parameters have to be taken into account
for $T\gg\Delta$ to explain the origin of the maximum in $K$ vs $T$ at $\approx 0.1\Theta$.

In this context, it appears natural to seek an interpretation in more
conventional terms. The maximum in $K$ versus temperature indicates an increasing
magnetic rigidity of the kagom\'{e} bi-layer spin network. As in all the $d=2$
AF systems, we suggest that the maximum is an experimental signature of a
small reinforcement of the magnetic correlations. The quasi-constancy of $K$
for all $p$ below this maximum (only a slight decrease is observed) is
therefore an indication that the correlations are short ranged and
concomitantly do not depend on the dilution.

Let us recall the typical characteristics evidenced in the susceptibility of
$d=2$ AF networks with Heisenberg spins as reviewed in Ref.~\onlinecite{deJongh}.
At high-$T$, the $d=2$ susceptibility follows a Curie-Weiss law, but exhibits
a maximum in the susceptibility at a temperature $T_{max}\approx\Theta $, which is a consequence of
the development of magnetic correlations. However, in contrast with $d=3$ systems
where a long range order is established at $T_{c}\approx \Theta $, a $d=2$ system has a finite
correlation length at $T_{max}\approx\Theta$ that eventually diverges when
$T\rightarrow 0$ K. The low dimensionality of the lattice inhibits the full
development of the critical fluctuations. In the case of the kagom\'{e} bi-layer
of SCGO, we suggest that the geometric frustration is even more efficient in
preventing correlations to develop, moving the maximum in $\chi _{frustr}$
down to $T_{max}\approx 0.1\Theta $.

On the experimental side, new neutron diffraction measurements further
support this point of view. As mentioned in Sec.~\ref{intro}, neutron studies on SCGO
have established that the magnetic correlations in the SG phase ($T<T_{g}$)
are short ranged. C. Mondelli \textit{et al.} observed more recently the
neutron diffraction spectrum in a wider temperature range than the previous
experiments ($1.5$ K$\leq T\leq 200$ K).\cite{Claudia} As shown in Ref.~\onlinecite{Claudia},
the characteristic diffuse peak typical for short range
correlations ($\xi \approx 2\times d_{Cr-Cr}$) develops at $T\leq 60$ K,
precisely in the same temperature range where the maximum in $\chi _{frustr}$
occurs. This maximum is therefore the signature in the susceptibility of the
development of these short range correlations. The feeble decrease in
$\chi _{frustr}$ we observe simply reflects the finite value of $\xi$ at
$T<T_{max}$.

A recent quantum mean field theory study also points at this direction.\cite{Garcia}
The basic idea here is to explicitly account for the magnetic
correlations by constructing the ground state not on interacting spins, but on
interacting triangles or tetrahedras of spins. Within this framework,
Garc\'{\i}a-Adeva \textit{et al.} calculated the susceptibility of the kagom\'e and the
pyrochlore networks for various spin values, providing a reference for
comparisons to our Ga-NMR shift $K$ (their mean-field theory does not account
for the existence of a spatially non uniform susceptibility responsible for the NMR
linewidth broadening, see Sec.~\ref{section:dilution effects}).
In a recent paper,\cite{Garcia2} they also studied the influence of the
lattice dilution on the susceptibility of these networks.
This theory seems to capture most of the main features of our data:
the calculated susceptibility for the diluted kagom\'e and pyrochlore
networks exhibits a maximum at $T_{max}\sim 0.15\Theta$
followed by a feeble decrease of the susceptibility in the $0.05\Theta -0.15\Theta$
temperature range. Although the order of magnitudes found are very encouraging,
as we underline in the bottom panel of Fig.~\ref{fig:11}, more experimental and
theoretical studies are needed here to completely validate this model.
Indeed, Garc\'{\i}a-Adeva \textit{et al.} predict that the susceptibility's maximum
should be progressively washed out with increasing dilution and
eventually disappear (in a $S=5/2$ kagom\'e lattice for a $\sim 10-15\%$ dilution
and in a $S=3/2$ pyrochlore lattice for a $\sim 20-30\%$ dilution), a feature
that somewhat contradicts our observations. Furthermore, as evidenced in the
bottom panel of Fig.~\ref{fig:11}, the $T$-dependence of $K$ for $T>T_{max}$ is dilution-independent.
This would only agree with the susceptibility calculated for a diluted pyrochlore lattice, but
for $T<T_{max}$ the more pronounced decrease in $K$ we observe is opposite to what
is predicted for the pyrochlore lattice. Unfortunately the Ga-NMR intensity loss at
$T<15$ K (Sec.~\ref{exp data}) does not enable us to further verify their
model, in particular to verify whether there is a Curie-like upturn in the susceptibility
for $T\ll 0.05\Theta$. We conclude then that for an adequate comparison between data and theory, the
influence of the kagom\'e bi-layer structure should be taken into account in the
calculations. Also, the non random distribution of the non magnetic vacancies in
SCGO's kagom\'e bi-layer might induce quantitative differences between the calculated and experimental
susceptibilities, as pointed out in Ref.~\onlinecite{Garcia2}.

The general picture drawn by our overview
of the experimental and theoretical data is that even short range spin-spin
correlations clearly play a major role in SCGO, or more generally in all geometrically frustrated compounds.
Future studies are required in this direction to further uncover the low-$T$ properties of these systems.

\begin{table*}[t]
\caption{The parameters of the two component fit to $\chi _{macro}$
described in the text. At low dilution, the fit yields a stronger error.}
\label{tab:4}
\begin{ruledtabular}
\begin{tabular}{@{\extracolsep{0ptplus1fil}}cccccc}
 $p$  & $C$ (emu K/mol) & $\Theta$ (K) & $C_{def}$ (emu K/mol) & $\Theta _{def}$ (K) & $C_{def}/C$ ($\%$) \\
\colrule
$0.72$ & $14.0(2)$       & $467(19)$    & $0.53(3)$             & $2.3(1.0)$          &  $3.8(2)$           \\
$0.81$ & $17.6(3)$       & $650(15)$    & $0.39(3)$             & $0.9(1.0)$          &  $2.2(2)$           \\
$0.89$ & $20.5(4)$       & $752(17)$    & $0.34(3)$             & $0.6(1.0)$          &  $1.7(2)$           \\
$0.90$ & $22.0(6)$       & $792(25)$    & $0.24(3)$             & $0.5(2.0)$          &  $1.0(2)$           \\
$0.95$ & $23.5(7)$       & $826(22)$    & $0.15(3)$             & $0.2(2.0)$          &  $0.6(2)$           \\
\end{tabular}
\end{ruledtabular}
\end{table*}

\section{Dilution effects}
\label{section:dilution effects}

\begin{figure}[b]
\includegraphics[width=7cm,bbllx=50,bblly=275,bburx=485,bbury=700,clip=]{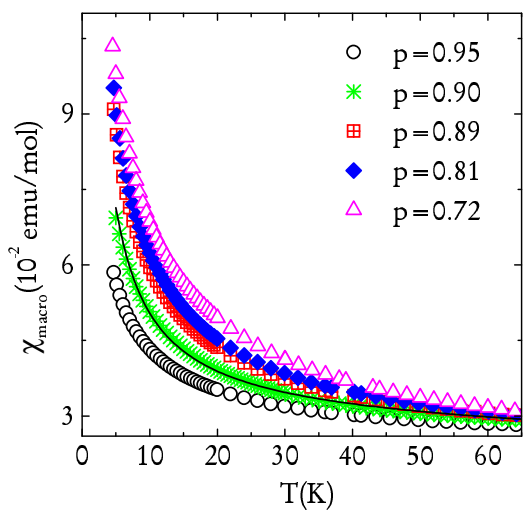}
\caption{The low-$T$ macroscopic susceptibility of some of the SCGO samples studied. The
solid line is the two component fit (Eq.(\ref{eqn:twocomp})) to the $p=0.90$
sample.}
\label{fig:14}
\end{figure}

In this section, we focus on the dilution-dependent susceptibility revealed
by the NMR linewidth at low-$T$. We show that the low-$T$ paramagnetic
behavior observed in $\chi _{macro}$ originates from the dilution of the
kagom\'{e} bi-layer. This feature is observed in nearly all the geometrically
frustrated AF compounds and appears as a generic behavior of this class of
materials.\cite{Schiffer}

Naturally, the macroscopic susceptibility could allow, in some way, to
appreciate the dilution effects through its low-$T$ behavior. Though, the
multiplicity of the Cr sites in SCGO complicates the analysis. The Ga($4f$)-NMR
width not only enables to evidence that the vacancy of a spin on the
network, i.e. the dilution, generates a paramagnetic defect, but also
allows to shed light on the more fundamental question concerning the nature
of the defect.

\subsection{Low-$T$ macroscopic susceptibility}

In Fig.~\ref{fig:14} we present the low-$T$ macroscopic susceptibility of
most of the samples studied. The low-$T$ paramagnetic upturn of $\chi
_{macro}$ is observed at all the concentrations, increasing in an appreciable
way with growing dilution. This establishes that the origin of this
contribution is related to the Ga/Cr substitution. The non magnetic gallium
vacancy on the network must therefore induce a perturbation which affects the neighboring
magnetic Cr$^{3+}$ ions. A paramagnetic defect is then generated, whose susceptibility $\chi_{def}$
is reflected in the low-$T$ behavior of $\chi _{macro}$.

In order to determine the exact low-$T$ dependence of the susceptibility
associated with the defects, the contribution to $\chi _{macro}$ of both
$\chi _{frustr}$ and $\chi _{pair}$ should in principle not be neglected.
Since $\chi _{frustr}$ cannot be determined through NMR below 15 K, we
follow Ref.~\onlinecite{Schiffer} and conveniently fit $\chi _{macro}$ by a
two component expression
\begin{equation}
\chi _{macro}=\frac{C}{T+\Theta }+\frac{C_{def}}{T+\Theta _{def}}
\label{eqn:twocomp}
\end{equation}
Anyhow, the corrections from $\chi _{frustr}$ and $\chi _{pair}$ are small
at low-$T$ compared to $\chi _{def}$ and do not affect significantly the
analysis. The first Curie-Weiss term roughly takes into account the
contribution from $\chi _{frustr}$ and $\chi _{pair}$, which dominate at
high-$T$. The second term, the more relevant for this section, also Curie-Weiss,
quantifies the contribution of $\chi _{def}$.

The fits, shown by a solid line in Fig.~\ref{fig:14} for the $p=0.90$ sample,
were performed in the $5$ K $\leq T\leq 350$ K range for all the samples and
reproduce correctly the low-$T$ behavior of $\chi _{macro}$. The fitting
parameters are presented in Tab.~\ref{tab:4} and are in agreement with
Ref.~\onlinecite{Schiffer}. The values of $C$ and $\Theta $ yield an overestimate
of $\approx 10-20\%$ compared to those of Tab.~\ref{tab:1}. The values of $C_{def}$
are in agreement with the analysis of Sec.~\ref{section:kagome bi-layer susceptibility}. As expected,
$C_{def}$ decreases with the dilution and most importantly is found
to be only a weak fraction of the constant $C$, $C_{def}/C\sim 1\%$ (Tab~\ref
{tab:4}). Finally, the Curie-Weiss temperature $\Theta _{def}$, which is an
indication of the average interaction between the defects, is negligibly small
within error bars except for the $p=0.72$ sample for which it is at most $2$
K.

\begin{figure}[b]
\includegraphics[width=6.5cm,bbllx=30,bblly=250,bburx=520,bbury=705,clip=]{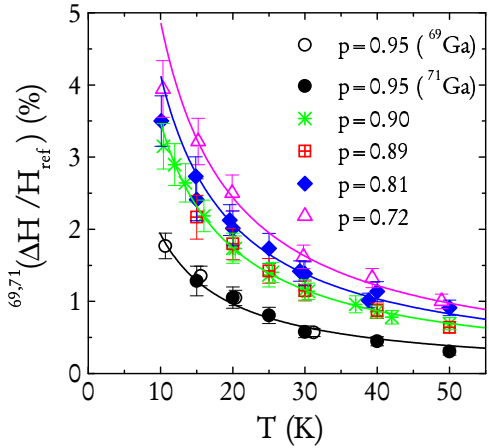}
\caption{$^{69,71}\Delta H$ versus $T$ for some samples studied.
$^{69,71}\Delta H$ is extracted from the $\nu _{rf}=40.454$ MHz
spectra following the fit described in Sec.~\ref{line analysis}. $^{69,71}\Delta H$ is
normalized by the reference field $^{69,71}H_{ref}$ to superimpose results
from the two isotopes (as explicitly shown for $p=0.95$). The solid lines are
the Curie-Weiss fits $C_{def/NMR}/(T+\Theta _{def/NMR})$.}
\label{fig:15}
\end{figure}

Strictly speaking, it should be finally noticed that $\chi _{macro}$ is not
sufficient on its own to establish experimentally whether the defects stem
from the dilution of the kagom\'{e} bi-layer. Indeed, the Ga/Cr substitution
on a Cr($4f_{vi}$) site can in principle break a spin pair and free a
paramagnetic spin, that can then contribute to the paramagnetic upturn of
$\chi _{macro}$. In contrast with $\chi _{macro}$, Ga($4f$)-NMR is not sensitive to
Cr($4f_{vi} $) and, as presented in the next subsection, allows a better
understanding of the dilution effects in SCGO.

\subsection{Dilution effects through the Ga(4f)-NMR width}
\label{Dilution effects through the Ga(4f)-NMR width}

\begin{figure}[t]
\includegraphics[width=8.6cm,bbllx=120,bblly=280,bburx=495,bbury=740,clip=]{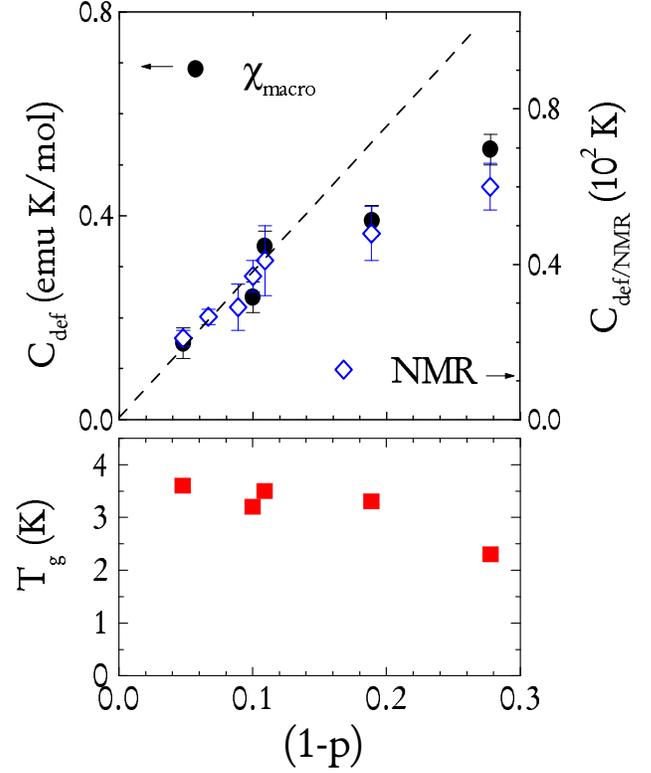}
\caption{\textit{Top panel}: $C_{def}$ extracted from $\chi _{macro}$ (\textit{left})
and $C_{def/NMR}$ (\textit{right}) extracted from the NMR width $^{69,71}\Delta H$
versus dilution $(1-p)$. The dashed line is a guide to the eye. \textit{Bottom panel}: The
SG temperature $T_{g}$ of our SCGO samples versus dilution.}
\label{fig:16}
\end{figure}

In Fig.~\ref{fig:15} we present the $T$-dependence ($10$ K$\leq T\leq 60$ K)
of the low-$T$ width of $^{69,71}$Ga($4f$) ($^{69,71}\Delta H$) for five of
the seven samples studied. We recover here the properties mentioned about
the raw spectra: $^{69,71}\Delta H$ increases as the temperature drops
and is very sensitive to the dilution, in contrast with $K$. The low-$T$ behavior of
$^{69,71}\Delta H$ bears strong similarities to the one of $\chi _{macro}$.
The perfect scaling of the two isotopes widths $^{69,71}\Delta H$ normalized
by the reference field $^{69,71}H_{ref}$ ($=\nu _{rf}/^{69,71}\gamma $) $-$
explicitly shown for the $p=0.95$ sample in Fig.\ref{fig:15} $-$ underlines
the magnetic origin of the low-$T$ broadening ($^{69,71}\Delta H\propto
^{69,71}H_{ref}$), in agreement with the prior study on the $p=0.90$ sample.
\cite{Philippe}

The fact that the $^{69,71}$Ga($4f$) linewidth increases rapidly at low-$T$
and with the dilution, whereas the shift varies little, establishes the
existence of a susceptibility that is \textit{spatially inhomogeneous} over
the kagom\'{e} bi-layer, due to the defects associated with the Ga/Cr substitution.
Since the Ga($4f$) nuclei are only coupled to the kagom\'{e} bi-layer, the
\textit{defects} probed by Ga-NMR are necessarily \textit{localized in the
kagom\'e bi-layer}.

\begin{figure}[t]
\includegraphics[width=7cm,bbllx=50,bblly=285,bburx=445,bbury=695,clip=]{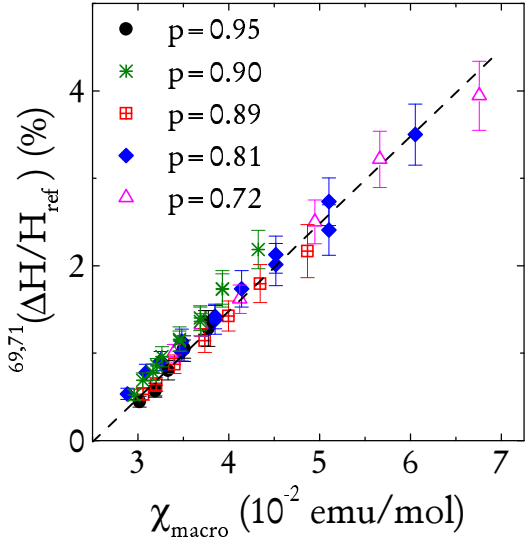}
\caption{$^{69,71}\Delta H$ versus $\chi _{macro}$ for some of the samples
studied ($T$ is an implicit parameter).}
\label{fig:17}
\end{figure}

The effect associated to the dilution is not reduced to a mere suppression of a
Cr site without any magnetic signature over the neighboring spins (i.e.
a magnetic defect) as indicated by the large modification of the $^{69,71}$Ga($4f$)
lineshape at low-$T$. Indeed, in the opposite case, one would expect a
lineshape which results from the gallium nuclear spins coupled to various
Cr-environments all bearing the same susceptibility. One would therefore
expect the linewidth to scale with the susceptibility, then $K$, which
varies only by 20\% from 60 K to 15 K, whereas the linewidth increases by one
order of magnitude. For this very same reason, a spatial distribution of the
hyperfine constant, which yields also a width $\propto K$, cannot justify the
broadening observed and also has to be ruled out.

The $T$-dependence of $^{69,71}\Delta H$ is correctly reproduced by a
Curie-Weiss law of the form $C_{def/NMR}/(T+\Theta _{def/NMR})$, where $%
C_{def/NMR}$ is a constant and $\Theta _{def/NMR}$ a Curie-Weiss
temperature. The fits are presented in Fig.~\ref{fig:15} as solid lines. The
values of $\Theta _{def/NMR}$ are extremely low, at maximum of $\approx 4$ K
and vary \textit{randomly} with dilution. Although we cannot exclude
completely a Curie-Weiss behavior, especially at high dilution, we may
consider the evolution in temperature of $^{69,71}\Delta H$ as very close to
a Curie law. The values of $C_{def/NMR}$ are plotted as a function of $(1-p)$
in the top panel of Fig.~\ref{fig:16} (right y-axis) along with the values of $C_{def}$
extracted from $\chi _{macro}$ by Eq.(\ref{eqn:twocomp}) (left y-axis).\cite
{average_p} As shown, the dependencies of both $C_{def}$ and $C_{def/NMR}$ on ($1-p$) perfectly match,
suggesting that $^{69,71}\Delta H$ and $\chi _{macro}$ probe the same
susceptibility at low-$T$, i.e. $\chi _{def}$. This is further underlined in
Fig.~\ref{fig:17} by the linear relationship between $^{69,71}\Delta H/H_{ref}$
and $\chi _{macro}$ at low-$T$. Since the defects probed by Ga-NMR
are located in the kagom\'e bi-layer, Fig.~\ref{fig:16} and Fig.~\ref{fig:17}
indicate that the defects on the Cr($4f_{vi}$) sites yield a contribution
scaling with the one from the kagom\'e bi-layer or little contribute to
$\chi _{macro}$. In view of the quantitative analysis of $C_{def}$ presented in
Ref.~\onlinecite{para4f}, the latter explanation is the more plausible. Finally, we note
in Fig.~\ref{fig:16} that $C_{def/NMR}$ and $C_{def}$ are linear with $(1-p)$ at low dilutions ($p>0.90$), and
become progressively sub-linear with ($1-p$) at higher dilutions. This deviation is directly
perceptible on the spectra of Fig.~\ref{fig:9}, where the FWHM of the $p=0.81$ sample is clearly
not $\approx 4$ times that of the $p=0.95$ sample.

\subsection{Discussion on the nature of the defect}

\subsubsection{Models for the localized defects}

To our knowledge, two models describe the dilution effects in a geometrically
frustrated system. Both are constructed in a classical approach where the
ground state corresponds to a minimization of the exchange energies on
triangles (or tetrahedras) of spins.

The most recent model was proposed by Moessner
\textit{et al}.\cite{Moessner} The basic idea here is that a paramagnetic
moment in the kagom\'{e} network is generated when two vacancies are adjacent on a
triangle (3 adjacent vacancies are needed for the pyrochlore network). The
defect corresponds in this case to a unique spin of the network, and at low
dilution the defect's susceptibility is $\chi _{def}\sim (1-p)^{2}/T$ for the
kagom\'{e} network ($\sim (1-p)^{3}/T$ for the pyrochlore network). This
prediction is in striking contrast with the linear variation of the NMR
width on dilution reported in Fig.~\ref{fig:16}. To reconcile our data to
the theory of Ref.~\onlinecite{Moessner}, it was suggested in Ref.~\onlinecite{Henley}
that the magnetic broadening of the width may reflect a distribution of random local fields
($\propto \chi _{def}$), due to the coupling of each gallium nucleus to the 12
neighboring Cr sites of the kagom\'e bi-layer. Basically, the NMR width $\Delta H$
would then probe the mean deviation of the susceptibility on these 12 sites
(and not simply $\chi _{def}$), thus $\Delta H\propto\sqrt{(1-p)^{2}}/T$.
Although appealing, this scenario is quite unlikely since the
macroscopic susceptibility measurements which are insensitive to the effect
of a local summation of fields, do not yield a quadratic $(1-p)^{2}$
variation on dilution.

\begin{figure}[b]
\includegraphics[width=7cm,bbllx=65,bblly=275,bburx=520,bbury=700,clip=]{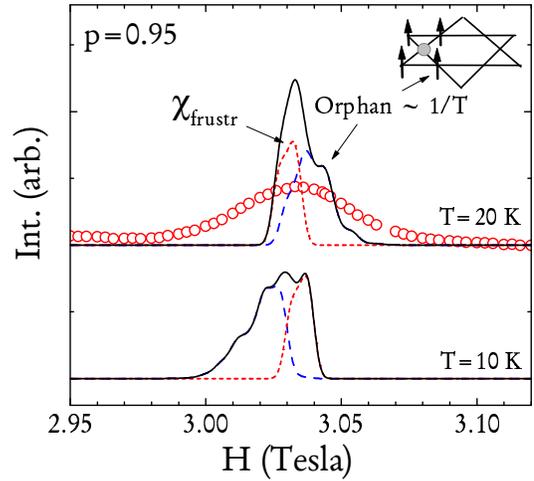}
\caption{The simulated sweep field central line of $^{71}$Ga($4f$) at $T=20$
K and $T=10$ K ($\nu _{rf}=40.454 MHz$) if orphan spins n.n. to the vacancy were present in the kagom\'{e} bi-layer
of the $p=0.95$ sample. Also shown is the experimental central line of $^{71}$Ga($4f$) recorded
for $p=0.95$ at 20 K (open circles). The area of the experimental and of the simulated lines are normalized.}
\label{fig:18}
\end{figure}

The second model $-$ purely empirical $-$ was proposed in Ref.~%
\onlinecite{Schiffer}. There, a defect is generated by a vacancy which
perturbs the neighboring spins, uncorrelating them from the rest of the spin
network. These uncorrelated spins were baptized ``orphan'' spins. Two
distinct spin populations then coexist: the orphan spins with a paramagnetic
behavior making up the defect, and the frustrated spins outside the defect
with a susceptibility equal to $\chi _{frustr}$. In order to examine the
implications of the orphan model on the Ga($4f$)-NMR low-$T$ spectrum, we
assume, as a starting point, that the defect is built only on the orphan spins
nearest neighbors (n.n.) of the vacancy (Fig.~\ref{fig:18}). We consider the
important case of the $p=0.95$ sample where the dilution is weak enough to
allow refined conclusions and where the Ga/Cr substitution occurs only on
the kagom\'{e} layers (the Cr($2a$) site is not substituted, see Fig.~\ref
{fig:3}). A simulation allows us to model the expected lineshape.
To do so, we start by randomly suppressing $5\%$ of the spins on
the kagom\'{e} network to account for the existence of non magnetic
vacancies. The n.n. spins of the vacancies are given a susceptibility $\propto \mu
_{eff}^{2}(def)/T$ where $\mu _{eff}(def)\approx 1.5\mu _{B}$ (evaluated
later in the section) to account for the orphan spins, whereas the remaining
Cr sites a susceptibility $\chi _{frustr}$. Next, we classify the various
nuclear populations along their Cr-environments. To construct the
spectrum, each population is associated to a quadrupole line simulated with
the parameters of Tab.~\ref{tab:2}, with a shift reflecting their
Cr-environment and an intensity weighted by the population size. Finally,
the Ga(4f)-NMR simulated line is obtained by summing the spectra associated
to all the nuclear populations. The line simulation for $T=20$ K and $T=10$ K is
presented in Fig.~\ref{fig:18}. As shown, neither the broadening, nor the
simulated lineshape agree with our $^{71}$Ga($4f$) low-$T$ experimental
spectrum. Indeed, in the $p=0.95$ sample, $\approx 35\%$ of the gallium
nuclei do not probe an orphan spin, hence $\approx 35\%$ of the spectral
weight yields a shift $\propto \chi _{frustr}$ (short dashed line in Fig.~\ref
{fig:18}). The majority of the gallium nuclei probe at least one orphan
spin, so that $\approx 65\%$ of the spectral weight yields a shift dominated
by a $1/T$ contribution at low-$T$ (dashed line in Fig.~\ref{fig:18}).
\textit{Independently on the value of} $\mu _{eff}$, one would therefore
expect to observe $\approx 65\%$ of the spectral weight to shift toward the
low fields with decreasing temperature whereas the remaining part is little
shifted. This would yield a non-observed asymmetric broadening of the NMR
line with decreasing temperature.

Clearly, a model built on orphan spins n.n. of the vacancy is unsatisfactory
and the perturbation generated by the vacancy must be extended in space. Can
we then imagine the defect to be made up of orphan spins which are first,
second, third etc. neighbors of the vacancy? Since the spin-spin
interaction is AF this situation is very unlikely to occur.\cite{orphan}
As we show below, an extended staggered perturbation constructed on an AF
interaction is conceivable.

\subsubsection{A defect built on AF correlations}

In Sec.~\ref{section:kagome bi-layer susceptibility} we showed via the
observed NMR shift that $\chi _{frustr}$ exhibits a maximum in
temperature, a common feature for the AF correlated systems. It is therefore
natural to seek an interpretation for the origin of the paramagnetic defect
also in the general context of the AF correlated systems. Indeed in the AF systems
such as the $d=1$ spin chains,\cite{spinchains} the quasi$-d=2$ spin ladders,\cite
{spinladders} and the $d=2$ cuprates,\cite{cuprates} it is now well established
that a vacancy (or a magnetic impurity) generates a long range oscillating
magnetic perturbation and creates a paramagnetic component in the
macroscopic susceptibility. The \textit{symmetric} broadening of the NMR
line observed in these systems is to be related to the \textit{oscillating} character of the
perturbation. We propose that the dilution effects of the SCGO kagom\'{e}
bi-layer can be described by the same physics of these correlated systems.

We recall first some basics concerning this model. The presence of a vacancy
in the AF correlated network of spins develops a magnetic perturbation,
which, in a general fashion, affects a spin at the lattice position $\roarrow r$ by
\[
m(\roarrow r)\sim \chi ^{\prime }(\roarrow r)H_{ref}
\]
where $m(\roarrow r)$, $\chi ^{\prime }(\roarrow r)$ and $H_{ref}$ are the
magnetization, the susceptibility and the applied magnetic field (constant
over $\roarrow r$). For an AF system, $\chi ^{\prime }(\roarrow r)$ is
peaked at a given vector $\roarrow Q$. As an example, we choose $\chi
^{\prime }(\roarrow r)$ to have a gaussian shape
\[
\chi ^{\prime }(\roarrow r)=\chi ^{\ast }\mathcal{G}(\roarrow r\cdot
\roarrow
Q)exp(-r^{2}/4\xi ^{2})
\]
where the function $\mathcal{G}(\roarrow r\cdot \roarrow Q)$ is oscillating
and periodic in $\roarrow r$ with a periodicity related to $\roarrow
Q$. $\xi$ is the spin-spin correlation length and $\chi ^{\ast }$ the
amplitude of $\chi ^{\prime }(\roarrow r)$. This translates into an
oscillating polarization of the spin network with a periodicity of $\sim
Q^{-1}$ and damped over $\sim \xi $.

\begin{figure}[b]
\includegraphics[width=8.6cm,bbllx=130,bblly=200,bburx=500,bbury=570,clip=]{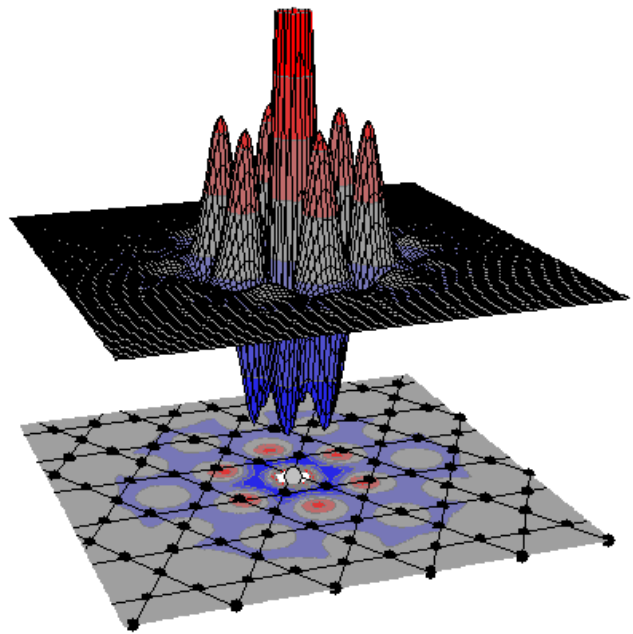}
\caption{A gaussian staggered polarization of arbitrary intensity generated by
a vacancy (center of the figure) on a 50 site kagom\'e lattice. $\xi$
is twice the spin-spin distance and the staggered vector is $\roarrow Q=(1/3,1/3)$,
i.e. the vector of the $\sqrt{3}\times\sqrt{3}$ kagom\'e spin configuration.}
\label{fig:19}
\end{figure}

In Fig.~\ref{fig:19} we illustrate this with a polarization calculated for
the kagom\'{e} network. In the absence of low diluted single crystals
necessary to determine the exact form of $\chi ^{\prime }(\roarrow r)$ and
the nature of the vector $\roarrow Q$ through neutron diffraction
experiments, we assume a gaussian shape for $\chi ^{\prime}(\roarrow r)$ and
take the reciprocal vector $\roarrow Q=(1/3,1/3)$ of the so-called
kagom\'{e} $\sqrt{3}\times \sqrt{3}$ spin configuration. By selecting this
vector, we partially account for the effects related to frustration.

The staggered shape of the polarization ensures that the NMR line of the nuclei
probing the spins of the lattice is broadened on both sides, i.e.
that the broadening is symmetric and scales with the applied field, in
agreement with our data. \textit{A non staggered polarization would generate
only local fields pointing in the same direction and the NMR line would
then be only broadened on one side, ie the broadening would be asymmetric (as
in the orphan spin model)}. The actual shape of the broadening (gaussian,
lorentzian etc.) depends on $\chi ^{\prime }(\roarrow r)$, on the vector
$\roarrow Q$ and on the geometrical details of the nucleus-spin coupling.

On the macroscopic stand, the vacancy generates a paramagnetic defect with a
macroscopic susceptibility
\begin{equation}
\chi _{def}=\frac{(1-p)\mu _{eff}^{2}(def)}{3k_{B}T}  \label{eqn:susdef}
\end{equation}
where $\mu _{eff}(def)$ is the effective moment of a defect. Although the
exact nature of the defect is not well established and is still the object of
theoretical studies,\cite{theorydefect} two scenarios may be conceived. 1)
The defect is the sum of all the polarizations, in other words the magnetization
$M=\sum_{\roarrow r}^{{}}{m(\roarrow r})$, whose susceptibility is
paramagnetic. The $1/T$ variation of the overall moment has then no
intuitive explanation. 2) The vacancy generates a localized paramagnetic
moment on the n.n. spins, which generates the staggered response $\chi
^{\prime }(\roarrow r)$ in the spin network. In this case, one needs to
understand microscopically why a paramagnetic moment is created.

Quantitatively, the expected NMR broadening is magnetic and related to $\chi
^{\ast }$ and $\xi $ by
\begin{equation}
\Delta H\propto \mathcal{A}\chi ^{\ast }\mathcal{F}(\xi ,1-p)\frac{H_{ref}}{T%
}  \label{eqn:width}
\end{equation}
where $\mathcal{A}$ is the hyperfine coupling constant between the spin and
the nucleus. $\mathcal{F}$ is a function of $\xi$ and of the dilution
$(1-p)$. The form of $\mathcal{F}$ depends on the form of $\chi ^{\prime }(\roarrow r)$.

The NMR width is sensitive to all variation in $T$ of $\xi (T)$ and $\chi ^{\ast }(T)$
in contrast with the macroscopic measurements. The surprising
simple $1/T$ variation of $\Delta H$ we observe by Ga-NMR in SCGO might
indicate that once the correlations set in below 50 K, very little changes
occur for $\xi (T)$ and $\chi ^{\ast }(T)$ due to the high frustration.
Concerning $\xi (T)$, neutron measurements are very rewarding since they do
infer that the spin-spin correlation is constant with temperature when
$T\leq 60$ K (Sec.~\ref{section:kagome bi-layer susceptibility}).

At low dilution, where the interaction between the defects can be neglected,
$\mathcal{F}(1-p)\propto (1-p)$. The evolution of $C_{def/NMR}$ with the dilution
reflects the one of $\mathcal{F}(\xi ,1-p)$. Fig.~\ref{fig:16} indicates
that $C_{def/NMR}$ is linear with $(1-p)$ at low dilution, but deviates
progressively from linearity at higher dilution. This deviation is indeed
expected. It can be associated with an ``interference'' between the
polarizations induced by the nearby vacancies. This phenomenon, known now for
several years, was first evidenced in the case of the RRKY polarization in
the dilute alloys.\cite{alloys} It is very important to note that $C_{def/NMR}$
likely extrapolates to 0 when $p\rightarrow 1$ or in the worst case to a
value lower than the observed values of $C_{def/NMR}$ for $p\leq 0.95$.
Since the SG state occurs at nearly the same temperature in all the samples
and $T_{g}$ varies the opposite way of $1-p$ (bottom panel of Fig.~\ref{fig:16}), this clearly demonstrates that
\textit{the origin of the SG state is not related to the dilution-induced
defects}. Whether a small residual $p$-independent Curie term associated
with intrinsic (topological) defects might explain the freezing is still a
matter of speculation as higher Cr-concentrations than $p>0.95$ would be
necessary to conclude.

To summarize, our NMR and macroscopic susceptibility results agree perfectly
with a picture of a defect built on the AF correlations of the frustrated
network. In conclusion we now elaborate on the effective value of the
paramagnetic moment detected in the low-$T$ macroscopic susceptibility.

Following Eq.(\ref{eqn:susdef}) and from the $C_{def}$ constant of Tab.~\ref
{tab:4}, we deduce a nearly $p$-independent value for the defect effective
magnetic moment of $\mu _{eff}(def)\approx 1.5\mu _{B}$ smaller than the
expected 3.87 $\mu _{B}$ value for a free $S=3/2$ spin. The weak ratio $%
C_{def}/C$ of Tab.~\ref{tab:4} is then simply explained in this viewpoint
since $C_{def}/C=(1-p)\mu _{eff}^{2}(def)/p\mu _{eff}^{2}(Cr^{3+})\approx
0.15(1-p)/p$. We can either speculate that the geometric frustration might
diminish the efficiency of the polarization induced by the vacancy,
indication that the nature of the defect is complex or the paramagnetic
defect can be modeled into a $S=1/2$ spin quantum state. Also, in the case
where $\mu _{eff}(def)$ results from the sum of the oscillating spin
polarization there is no reason to find a $3/2$ value. Further theoretical
work in this direction is indeed required to model the defect and achieve a
quantitative understanding of our data.

\section{Summary and concluding remarks}

\label{Conclusion}

The SCGO kagom\'{e} bi-layer compound was studied for the first time through
a local probe over a wide variety of Cr-concentrations. New and original
properties were unravelled for the archetype of geometrically frustrated
compounds.

Ga($4f$)-NMR allowed us to observe the kagom\'{e} bi-layer susceptibility
$\chi _{frustr}$, otherwise non accessible by macroscopic measurements. A
maximum in $\chi _{frustr}$ occurs at a  temperature ${T_{max}\approx
0.1\Theta }$, robust to a dilution as high as $\approx 20\%$. No gaped
feature is evidenced. The maximum signals the appearance of short ranged
magnetic correlations in the kagom\'{e} bi-layer, as observed in all the AF $d=2$
Heisenberg systems. $T_{max}$ is a new energy scale for SCGO, which sets a
new constraint on the theoretical models.

A close examination of the dilution effects by Ga($4f$)-NMR allowed us to
establish, on a firm experimental ground, that paramagnetic defects are
present on the kagom\'e bi-layer. The defects stem from the vacancies in the
spin network, i.e. from the substitution of magnetic Cr$^{3+}$ ions
with non magnetic Ga$^{3+}$ ions. The defects are responsible for the low-$T$
Curie upturn of $\chi _{macro}$. The macroscopic susceptibility therefore does not probe an
intrinsic property of the kagom\'e bi-layer at low-$T$, but mainly a property
related to the dilution. Interestingly, the SG transition is detected precisely
in the low-$T$ upturn of $\chi _{macro}$. As there are experimental
signatures suggesting that this transition is an intrinsic feature of SCGO,
\cite{Martinez2,ClaudiaTg} this means that the defects do not trigger the
freezing, but do freeze at $T<T_{g}$. The origin of the SG transition
remains to be elucidated.

Finally, we proposed a mechanism that justifies the existence of the
paramagnetic defects, which, to our knowledge, is the only one that gives a
consistent interpretation for both the Ga($4f$)-NMR and the macroscopic
susceptibility results. Our NMR data point at a defect which possesses
strong analogies with the defects observed in most of the AF correlated systems $-$
where the presence of a vacancy (or a magnetic impurity) generates a
staggered response from the spin network. Up to now, the dilution in SCGO, or
more generally in the frustrated systems, has been considered a parasitical
effect to be minimized in order to measure the intrinsic properties related
to the geometric frustration. In light of our results, it is interesting to
reverse the problematic. The study of the response from the magnetic lattice
to a non magnetic vacancy can be an indirect way of probing the intrinsic
properties of the frustrated network such as $\xi (T)$ or $\chi ^{*}(T)$.
The experimental study and the theoretical models to describe this response are
a new domain of investigation for the geometrically frustrated systems.

\begin{acknowledgments}
It is a pleasure to thank J. Bobroff, H. Alloul, A. Keren, C. Lhuillier and F. Mila
for fruitful discussions.
\end{acknowledgments}

\appendix

\section*{Appendix: susceptibility of an isolated spin pair}
\label{app}

We consider pairs of spins $s$ with an AF exchange constant $J$. Their states
are labeled by their total spin quantum number, $S=0,1,2,...,2s$, and lie
at energies $E_{S}=JS(S+1) / 2$ relative to the ground state. The partition
function for $N$ independent pairs ($2N$ being the total number of spins involved in
pair formation) is
\begin{eqnarray*}
\mathcal{Z}=\left[\sum_{S=0,2s}g_{S}exp(-\beta E_{S})\right] ^{N}
\end{eqnarray*}
where $g_{S}$ is the degeneracy of each level and $\beta =1/k_{B}T$. The sum
is taken over $S=0,1,2,3...2s$. We now apply an external field $H$ to
calculate the susceptibility. The degeneracy is lifted by a Zeeman splitting
$E_{m_{S}}(H)=g\mu _{B}Hm_{S}$ so that
\begin{eqnarray*}
\mathcal{Z}(H)&=&\left[ \sum \limits_{S=0,2s}\ \sum \limits_{m_{S}=-S,S} exp%
\left[-\beta (E_{S}+E_{m_{S}})\right]\right] ^{N} \\
\end{eqnarray*}
where m$_{S}=-2s,...,0,...,2s$ are the Zeeman levels for a spin $S$. From
the expression of $\mathcal{Z}(H)$ we then derive the susceptibility of $N$
isolated spin pairs
\begin{eqnarray*}
\chi =\frac{Ng^{2}\mu _{B}^{2}}{k_{B}T}\frac{\sum\limits_{S=0,2s}\sum%
\limits_{m_{S}=-S,S}m_{S}^{2}exp(-\beta E_{S})}{\sum%
\limits_{S=0,2s}g_{S}exp(-\beta E_{S})}
\end{eqnarray*}

The susceptibility of a Cr(4f$_{vi})$-Cr($4f_{vi}$) spin pair $(s=3/2)$ per
formula unit of SCGO is therefore
\begin{eqnarray*}
\chi _{pair}&=&p_{4f_{vi}}^{2}\frac{\chi}{N} \\
&=&\frac{p_{4f_{vi}}^{2}\mu _{eff}^{2}}{s(s+1)k_{B}T}\times \\
& &\frac{2exp(-\beta J)+10exp(-3\beta J)+28exp(-6\beta J)}{1+3exp(-\beta
J)+5exp(-3\beta J)+7exp(-6\beta J)} \\
\end{eqnarray*}
where we have introduced the effective moment $\mu _{eff}=\sqrt{g^{2}\mu
_{B}^{2}s(s+1)}$. The value of the AF exchange constant, i.e. of the
singlet-triplet gap, is $J=18.6(1)$ meV.\cite{Lee} To account for the dilution
effects, $\chi _{pair}$ is weighted by $p_{4f_{vi}}^{2}$, the statistical
probability of having a pair of Cr($4f_{vi}$)-Cr($4f_{vi}$) per formula unit
of SCGO. The value of $p_{4f_{vi}}$ is known from neutron refinements (Fig.~%
\ref{fig:3}). Note that in the limit $T\gg J/k_{B}$, we recover the sum of
two $s=3/2$ paramagnetic susceptibilities: $2p_{4f_{vi}}^{2}\mu
_{eff}^{2}/3k_{B}T$.




\begin{thebibliography}{99}
\bibitem{69Ga} The contribution of the non-stoichiometric gallium to the $^{69}$%
Ga spectrum is analogous to the $^{71}$Ga spectrum. However, $^{69}$Ga($sub$%
) is not resolved: the quadrupole effects, more important as compared to the
other isotope, enlarge both $^{69}$Ga($4f$) and $^{69}$Ga($sub$) lines and
result into two overlapping lines (see Fig.~\ref{fig:6}).
\bibitem{T2} Strictly speaking, with this procedure we also convolute the $%
4e$ line, which stays constant in temperature; however the $^{69,71}$Ga($4e$%
) line is sufficiently large compared to the $^{69,71}$Ga($4f$) line so that the
convolution has only a negligible effect on the real shape of its spectrum.
\bibitem{hyperfine} Since it is impossible to determine individually
$\mathcal{A}_{12k}$ and $\mathcal{A}_{2a}$, respectively the hyperfine constants
of the Ga($4f$)-O-Cr($12k$) and Ga($4f$)-O-Cr($2a$) couplings, we use an average
hyperfine constant  $\mathcal{A}=(9\mathcal{A}_{12k}+3\mathcal{A}_{2a})/12$.
A calculation in Ref.~\onlinecite{AmitNMR} suggests $\mathcal{A}_{2a}/\mathcal{A}_{12k}\approx 2$.
\bibitem{average_p} For clarity, we do not take into account the non random
character of the Ga/Cr substitution. Since each Ga($4f$) nucleus is coupled
to $9$ Cr(12k)-sites and only to $3$ Cr($2a$)-sites, which, moreover, are less
substituted than the Cr($12k$)-sites, it is safe to use an average $p$ equal
to the nominal concentration.
\bibitem{para4f} Taking a binomial distribution for Ga/Cr substitutions
on the $4f_{vi}$ site, the paramagnetic susceptibility resulting from the
broken spin-pairs would yield a Curie constant of $\approx 0.36$ emu K/mol for the
$p=0.95$ sample, more than twice the experimental value of Tab.~\ref{tab:4}. The
substitutions on the $4f_{vi}$ site are therefore sub-stoichiometric. Some
chemical constraint must favor substitutions by pairs of gallium ions,
minimizing the number of paramagnetic spins resulting from the broken
spin-pairs.
\bibitem{orphan} Furthermore, if this were the case, the gallium nuclei
in the highly diluted samples would be only coupled to paramagnetic orphan
spins. The gallium line would then shift as $1/T$ at low-$T$, in contrast with
our observations.
\end{thebibliography}
\end{document}